\documentclass[aps,prb,twocolumn,groupedaddress,showpacs,floatfix,eqsecnum]{revtex4}

\usepackage{graphicx}
\usepackage{latexsym}
\usepackage{amsmath,amssymb}
\usepackage{bm}
\usepackage{color}

\DeclareMathOperator{\Tr}{Tr}
\newcommand{\dvex}[1]{d^{2}\bm{\mathrm{#1}}}
\newcommand{\vex}[1]{\bm{\mathrm{#1}}}

\newcommand{\ord}[1]{\bm{\mathit{O}}\left(#1\right)}
\newcommand{\sigmah}{\hat{\bm{\sigma}}}
\newcommand{\sih}[1]{\hat{\sigma}^{#1}}
\newcommand{\kah}[1]{\hat{\kappa}_{#1}}

\newcommand{\tgu}{\widetilde{g_{u}}}
\newcommand{\tgun}{\widetilde{g_{u}}^{(0)}}
\newcommand{\bw}{\bar{w}}

\newcommand{\bGs}{\widetilde{\mathcal{G}}_{s}}
\newcommand{\bGt}{\widetilde{\mathcal{G}}_{\mathsf{T}}}
\newcommand{\gammab}{\bar{\gamma}}
\newcommand{\lambdab}{\bar{\lambda}}
\newcommand{\alphab}{\bar{\alpha}}
\newcommand{\betab}{\bar{\beta}}
\newcommand{\etab}{\bar{\eta}}
\newcommand{\xib}{\bar{\xi}}
\newcommand{\bg}[1]{\bar{g}_{#1}}
\newcommand{\dlambda}{\frac{d\phantom{\ln\Lambda}}{d \ln\Lambda}}

\newcommand{\be}{\begin{equation}}
\newcommand{\ee}{\end{equation}}

\newcommand{\reqs}[1]{Eqs.~(\ref{#1})}
\newcommand{\rref}[1]{(\ref{#1})}

\begin{document}

\title{Graphene via large N I: Renormalization}
\author{Matthew S. Foster}
\email{foster@phys.columbia.edu} 
\author{Igor L. Aleiner}
\affiliation{Department of Physics, Columbia University, New York, NY 10027}
\date{\today}

\begin{abstract}
We analyze the competing effects of moderate to strong Coulomb electron-electron
interactions and weak quenched disorder in graphene. 
Using a one-loop renormalization group calculation controlled within the large-$N$ 
approximation, we demonstrate that,
at successively lower energy (temperature or chemical potential) scales, 
a type of non-Abelian vector potential disorder always asserts itself as the dominant
elastic scattering mechanism for generic short-ranged microscopic defect distributions. 
Vector potential disorder is tied to both elastic lattice deformations (``ripples'') and 
topological lattice defects.
We 
identify several well-defined scaling regimes, 
for which we provide
scaling predictions for the electrical conductivity and thermopower,
valid when the inelastic lifetime due to interactions exceeds the elastic lifetime due to disorder.
Coulomb interaction effects should figure strongly into the physics of suspended graphene films,
where $r_{s} > 1$; we expect vector potential disorder to play an important role in 
the description of transport in such films.
\end{abstract}

\maketitle

\section{\label{Intro}INTRODUCTION}

The recent experimental realization of graphene,\cite{Geim1,Geim2,Kim} a single atomic monolayer of 
graphite, has refocused attention upon many 
fundamental questions regarding electronic transport in low dimensions.
Its remarkable bandstructure at zero doping has made graphene a candidate solid state analog for 
high energy particle phenomena, e.g.\ that described by the theories of quantum electrodynamics (QED) 
or chromodynamics (QCD).

Given the relative simplicity of graphene as solid state system, it seemed not unreasonable to hope
that a complete understanding of its basic electronic properties would emerge quickly. This expectation
has been undermined by several puzzling results in the first graphene experiments.\cite{Geim1,Geim2,Kim} 
Perhaps most paradoxical is the ``quasi-ballistic'' nature of its electronic transport: at zero magnetic field,
graphene's conductivity is a linear, particle-hole symmetric function of carrier density 
that varies little over a temperature range of several orders of magnitude,\cite{Geim1,Geim2,Kim} down to the 
lowest temperatures so far measured (30 mK).\cite{Kim2} 
At exactly zero doping, a condition termed the ``Dirac point,'' the conductivity assumes a minimum value 
of order the conductance quantum $e^{2}/h$; this ``minimum metallic'' conductivity varies only weakly between 
different samples possessing mobilities spanning an order of magnitude.\cite{Geim1,Geim2,Kim,Kim2}

One expects that any disorder present must be playing a crucial role in limiting the low-temperature conductivity
of undoped graphene. Potentially important sources of disorder in the experiments performed in 
Refs.~\onlinecite{Geim1,Geim2,Kim,MorozovNovoselov,Kim2} include remote charged 
impurities in the $\textrm{SiO}_{2}$ 
substrate,\cite{NomuraMacDonald1,NomuraMacDonald2,DasSarma1,DasSarma2,CheianovFalkoAltshulerAleiner} as well as corrugations or ``ripples'' 
in the graphene sheet.\cite{MorozovNovoselov,MeyerGeim,MorpurgoGuinea,FasolinoLosKatsnelson}
A semiclassical computation within the self-consistent Born approximation (SCBA)  
at the Dirac point gives a ``bare'' minimum metallic conductivity independent of
the disorder strength, and of order the conductance quantum,\cite{Fradkin,Lee,AndoSCBA,PeresGuineaCastroNeto}
in apparent agreement with experiment. Unfortunately, the SCBA is known to be inconsistent 
for massless Dirac fermions in 2D.\cite{NersesyanTsvelikWenger,AleinerEfetov} 
Das Sarma et al.\cite{DasSarma1,DasSarma2}\ have employed an alternative mean field treatment 
focused upon charged impurity scattering, and have concluded that the minimum metallic conductivity is in fact 
not universal, but should exhibit dependence upon the disorder distribution. 
Near zero doping, Cheianov et al.\cite{CheianovFalkoAltshulerAleiner} have further argued that percolation
effects become important; these authors used semiclassical percolation theory to derive a scaling form of the conductivity
in terms of the transmission coefficient between electron and hole puddles. 
Other recent 
analytical\cite{AleinerEfetov,OstrovskyGornyiMirlin1,MorpurgoGuinea,OstrovskyGornyiMirlin2,RyuMudryObuseFurusaki} 
and numerical\cite{NomuraMacDonald2,BardarsonTworzydloBrouwerBeenakker,NomuraKoshinoRyu}
work has focused upon the localization physics of massless Dirac electrons in the presence 
of various types of disorder. These and 
previous\cite{LFSG,NersesyanTsvelikWenger,CILogCFT,HatsugaiWenKohmoto,GuruswamyLeClairLudwig,
RyuHatsugai} studies have demonstrated that Dirac electrons may evade Anderson localization for 
certain fine-tuned disorder distributions, but the relevance of these results to graphene remains 
unclear. 

All of the references discussed above essentially ignore a potentially important aspect of graphene physics,
that of electron-electron interactions. 
Carriers native to the carbon sheet are ineffective at screening the long-ranged Coulomb potential for a 
system near zero doping. The dimensionless strength of the Coulomb 
interactions $r_{s} \sim e^{2}/\epsilon \hbar v_{F}$, with $v_{F}$ the Fermi velocity and 
$\epsilon$ the (effective) dielectric constant, is close to one in the 
experiments,\cite{Geim1,Geim2,Kim,MorozovNovoselov,Kim2} and should
exceed two for suspended films.\cite{MeyerGeim}
This is not large by the standards of bulk 3D metals. However, it is important to stress
that, near the Dirac point, Fermi liquid theory is \emph{not} expected to apply, at least in the absence
of disorder. Instead, the situation in graphene is analogous to that in the low-energy domains of 
QED and QCD in $3+1$ spacetime dimensions; the smallness of the fine structure constant allows for a consistent perturbative expansion
in QED for length scales of order or larger than the Compton wavelength. In the case of QCD, however, the 
coupling strength is not small at atomic distances, and in fact grows ever larger with increasing 
length scale. 
It is not apriori clear that the Coulomb interparticle interactions in graphene 
can be treated using perturbation theory in the coupling strength.

For weak Coulomb interactions and disorder, 
one can use 
the perturbative renormalization group (RG) to determine whether disorder or 
interaction effects should become 
dominant
at 
low temperatures. In this paper, we employ a large-$N$ generalization 
of the graphene field theory to treat the effects of moderate to strong Coulomb interactions; in physical graphene, 
$N = 4$ (see Sec.~\ref{Setup}, below). We study the resulting theory using a one-loop RG calculation perturbative 
in $1/N$ and in the disorder strength. 

The RG has been used many times before to study disorder effects in 2D Dirac 
fermions,\cite{LFSG,HatsugaiWenKohmoto,BERNARD,GuruswamyLeClairLudwig,AleinerEfetov} as well as the competition
between disorder and interaction effects in such 
systems.\cite{YeSachdev,Ye,StauberGuineaVozmediano,BDIFosterLudwig} 
Using a non-interacting model of graphene, Aleiner and Efetov\cite{AleinerEfetov} showed that a particular 
type of disorder, associated with long-wavelength potential fluctuations consistent with, e.g., 
screened
remote charged substrate impurities, dominates the flow to strong coupling in the absence of interactions. Stauber, Guinea, and 
Vozmediano\cite{StauberGuineaVozmediano} used the RG to treat both weak disorder and weak Coulomb interactions in graphene;
they recovered results previously obtained by Ye and Sachdev.\cite{YeSachdev,Ye} In particular, 
the weak-coupling RG for graphene demonstrates that a) in the absence of disorder, Coulomb interactions 
are irrelevant (in the sense of the RG); the non-interacting, clean Dirac description is \emph{stable}, 
and b) in the presence of generic disorder, the non-interacting, clean Dirac fixed point becomes \emph{unstable}, 
and the system flows toward strong disorder and interaction coupling. For weak Coulomb interactions, the flow 
to strong coupling is again dominated by scalar potential disorder, consistent with the formation of local 
electron and hole ``puddles.'' This picture would seem to suggest that the asymptotic low-energy physics in disordered
graphene should be the same as that in an ``ordinary,'' diffusive 2D electron gas 
(but see 
Refs.~\onlinecite{OstrovskyGornyiMirlin2,BardarsonTworzydloBrouwerBeenakker,NomuraKoshinoRyu,RyuMudryObuseFurusaki}). 
Multiple crossover regimes are possible depending upon assumptions regarding the microscopic disorder 
distribution.\cite{AleinerEfetov,OstrovskyGornyiMirlin1}

Our large-$N$ RG treatment of graphene follows closely a previous calculation of Ye,\cite{YeSachdev,Ye} who 
employed the same methodology to study strong Coulomb interaction and weak disorder effects in a Dirac system 
possessing a single Fermi point (generalized to $N$ flavors). 
Different from Ye, we treat the case of two inequivalent Fermi points (valleys); each valley 
is generalized to $N/2$ identical flavors. Our calculational framework therefore maintains the important distinction between disorder 
potentials that scatter particles between inequivalent valleys and those that do not. Further, we correct a 
calculational mistake made in Ref.~\onlinecite{Ye}, the correct determination of which crucially affects our results. 
(See Sec.~\ref{SLFNRG} and Appendix~\ref{APP-PsiSLFNRG} for details.)
Son\cite{Son} recently revisited the problem of Coulomb interactions at large-$N$ for clean Dirac electrons
in graphene; in the limit of vanishing disorder, we recover the findings of Ref.~\onlinecite{Son}.

We now summarize our primary results.
We show that, within the large $N$ approximation, 
disordered
graphene resembles QCD more than QED. The theory becomes
more strongly coupled at lower temperatures and longer length scales. 
This eventually leads to a breakdown of our RG calculation, so that we cannot 
determine the ground state phase within the framework employed here. However,
we are able to identify various scaling regimes which may play a role in 
future experimental observations, particularly in suspended graphene films.\cite{MeyerGeim}
Our main results are obtained via numerical integration of the RG flow equations, and are as follows:
a) The non-interacting, non-disordered Dirac fixed point is unstable upon the incorporation of generic, arbitrarily
weak disorder. b) For the case of weak ($r_{s} \ll 1$) or vanishing Coulomb interactions, 
the RG flows to strong scalar potential disorder, consistent with the electron-hole puddle 
picture.\cite{AleinerEfetov,DasSarma1,DasSarma2,CheianovFalkoAltshulerAleiner} 
c) Most importantly, for moderate ($r_{s} \sim 1$) to strong ($r_{s} \gg 1$) 
Coulomb interactions, scalar potential disorder fluctuations are 
\emph{parametrically} 
cut off via screening. The system 
flows to strong interaction and disorder coupling, but now, the runaway flow is dominated by different type of disorder, 
corresponding to a quenched SU(2) non-Abelian vector potential in the effective Dirac electron theory. 
This vector potential is 
implicated in ripples,\cite{MorozovNovoselov,MeyerGeim,OstrovskyGornyiMirlin1,MorpurgoGuinea} as well as  
the representation of topological lattice defects,\cite{TopologicalDisorder} 
in the low-energy theory.
Finally, we use the RG
to predict logarithmic temperature scaling in the dc conductivity within the Drude/Boltzmann transport regime,
and we also discuss thermal transport. 

The importance of quenched vector potential disorder in the context of graphene has been stressed
several times before in the literature.\cite{MorozovNovoselov,MeyerGeim,OstrovskyGornyiMirlin1,MorpurgoGuinea} 
All of these previous works neglected interparticle interactions, however, and overlooked the important fact that 
there is no physically plausible way of disordering a graphene sheet that would produce only non-Abelian 
vector potential disorder in isolation (see Sec.~\ref{LatticeModelandSym}, below). 
Our result establishes that, 
for moderate to strong Coulomb interactions in the large-$N$ approximation, non-Abelian vector potential 
disorder emerges \emph{generically} as the dominant source of randomness.
Finally, it is intriguing to note that the non-interacting graphene theory subject \emph{only} to non-Abelian 
vector potential disorder constitutes a very unusual, manifestly particle-hole symmetric quantum disorder 
model, for which several exact and/or non-perturbative results are 
known.\cite{NersesyanTsvelikWenger,CILogCFT,OstrovskyGornyiMirlin1} This non-Abelian Dirac disorder 
model is predicted to \emph{avoid} Anderson localization, possessing instead a perturbatively inaccessible, 
critical ground state,\cite{CILogCFT} a vanishing disorder-averaged, single-particle 
density of states,\cite{NersesyanTsvelikWenger} and a non-zero conductivity of order $e^{2}/h$, independent 
of the disorder strength.\cite{OstrovskyGornyiMirlin1} 

The outline of this paper is as follows.
In Sec.~\ref{Setup}, we write down the low energy field theory for disordered, interacting graphene, and 
generalize it to the case of large-$N$ in order to treat strong Coulomb interactions. We discuss the
microscopic interpretation of the various types of disorder that appear in the low-energy theory, 
and we review the quantum disorder universality classes obtained by fine-tuning this distribution.
In Sec.~\ref{PerturbativeCalc}, we perform a one-loop renormalization group calculation upon
the model defined in Sec.~\ref{Setup}. In Sec.~\ref{Results}, we present the results of this calculation, 
which consist of flow equations for the various parameters defining the graphene field theory.
We analyze these equations and attempt to gleam information about graphene's phase diagram
in an abstract disorder-interaction coupling constant space. 
Experimental consequences of our results in the form of scaling predictions for the dc conductivity 
and thermopower are summarized and discussed in Sec.~\ref{Physics}. 

A variety of elaborations and extensions are relegated to the Appendices.
Some technical details of the calculation presented in Sec.~\ref{PerturbativeCalc} are relegated to 
Appendix \ref{APP-PsiSLFNRG}, while a survey of several (unstable) fixed line structures 
appearing in the flow equations stated in Sec.~\ref{Results} is presented in Appendix~\ref{APP-FixedLines}. 
The scaling predictions enumerated in Sec.~\ref{Physics} are derived in Appendix~\ref{APP-Physics}.


\section{Model and setup\label{Setup}}

The plan of this section is as follows:
In subsection \ref{LatticeModelandSym}, we write down an effective field theory 
for graphene incorporating both quenched disorder and Coulomb interparticle interactions,
and we discuss (1) the role of discrete symmetries and (2) the different universality 
classes of disordered quantum systems that arise by fine-tuning the details of the
impurity or defect distribution. Most of this material is not new, see e.g.\ 
Refs.~\onlinecite{AleinerEfetov,OstrovskyGornyiMirlin1}; the purpose of this 
section is to provide a context for the results discussed in Sec.~\ref{Results}, 
and to establish notation. In subsection \ref{LargeN}, we construct a large-$N$ 
generalization of our graphene model, suitable for studying the effects of strong 
Coulomb interactions.

\subsection{Lattice model and effective field theory\label{LatticeModelandSym}}

The bandstructure of undoped graphene is well-approximated by the tight-binding
Hamiltonian
\begin{equation}\label{H0}
	H_{0} = - t \sum_{\langle \vex{r} \vex{r'} \rangle, \, s} c_{A s}^{\dagger}(\vex{r}) c_{B s}^{\phantom{\dagger}}(\vex{r'}) + \mathrm{H.c.},
\end{equation}
where $c_{A s}^{\dagger}(\vex{r})$ and $c_{B j s}^{\phantom{\dagger}}(\vex{r'})$ are creation and annihilation
operators for electrons on the $A$ and $B$ sublattices of the bipartite honeycomb lattice, respectively.
The index $s \in \{\uparrow,\downarrow\}$ 
denotes spin-$1/2$ components of the lattice electrons. The hopping amplitude $t$ in Eq.~(\ref{H0}) is 
purely real, and the sum on $\langle \vex{r} \vex{r'} \rangle$ runs over all nearest-neighbor $A-B$ bonds in the 
graphene sheet. (Further-neighbor hopping may be incorporated as a perturbation). Henceforth we adopt
units such that $\hbar = k_{B} = 1$.

A low-energy effective field theory for undoped graphene obtains by 
linearizing the bandstructure of Eq.~(\ref{H0}) in the vicinity of the 
two inequivalent $K$ and $K'$ 
Fermi points (also termed ``valleys''), where the particle-hole symmetric 
energy bands meet. Retaining only low-energy modes, we write
\begin{equation}\label{PsiDef1}
	c_{\sigma s}(\vex{r}) \sim 
	e^{i \vex{k_{F}}\cdot\vex{r}} \psi_{\sigma s}^{K}(\vex{r})
	+ e^{- i \vex{k_{F}}\cdot\vex{r}} \psi_{\sigma s}^{K'}(\vex{r}),
\end{equation}
where $\pm \vex{k_{F}}$ locates the $K$ and $K'$ points, respectively,
and $\sigma \in \{A,B\}$ labels the sublattice species.

We assemble the Dirac spinor
\begin{equation}\label{PsiDef2}
	\psi(\vex{k}) \equiv
	\begin{bmatrix}
	\phantom{-}\psi_{A}^{K}(\vex{k})\\
	\phantom{-}\psi_{B}^{K}(\vex{k})\\
	\phantom{-}\psi_{B}^{K'}(\vex{k})\\
	-\psi_{A}^{K'}(\vex{k})
	\end{bmatrix}.
\end{equation}
(The spin index $s$ has been suppressed in this equation.) In the low-energy
theory, we will consider states annihilated by $\psi(\vex{k})$ with momentum $k \leq \Lambda \ll k_{F}$,
where $\Lambda$ is a hard cutoff.

The Dirac spinor defined by Eq.~(\ref{PsiDef2}) is an eight-component object,
$\psi\rightarrow\psi_{a}$, with index $a \in \{1,\ldots,8\}$. The eight components arise
from the direct product of indices in the 2-dimensional sublattice $\{A,B\}$, valley 
$\{K,K'\}$, and spin $1/2$ $\{\uparrow,\downarrow\}$ component subspaces.
Later, we will generalize $\psi_{a}$ to $2 N$ components in order to perform an 
expansion in $1/N$. 
We introduce two commuting sets of Pauli matrices: 
the matrix $\sih{\alpha}$ acts in the sublattice $\{A,B\}$ space of Eq.~(\ref{PsiDef2}), while the matrix
$\kah{\beta}$ acts in the valley $\{K,K'\}$ space, with $\alpha,\beta \in\{1,2,3\}$.\cite{footnote-a}

Below, we write down the continuum field theory for graphene that incorporates the effects of both 
quenched disorder and Coulomb electron-electron interactions [see Eq.~(\ref{H})]. 
The clean model in Eq.~(\ref{H0}) possesses 
three crucial symmetries that might plausibly survive the incorporation of 
disorder: particle-hole symmetry (PH), time-reversal invariance (TRI), and spin SU(2) rotational symmetry. In the low-energy theory, 
the former are encoded in the operator level transformations
\begin{align}
	\psi(\vex{r}) &\rightarrow -\sih{1} \kah{1} 
	\left[
	\psi(\vex{r})^{\dagger}
	\right]^{\mathsf{T}},
	&&\textrm{(PH)};\label{PH}
	\\
	\psi(\vex{r}) &\rightarrow \sih{2} \kah{2} \,
	\psi(\vex{r}),
	&&\textrm{(TRI)}.\label{TRI}
\end{align}
In this equation, $\mathsf{T}$ denotes the ordinary (matrix) transpose.
The particle-hole (time-reversal) transformation defined by these equations is unitary (antiunitary).
[Schematically writing $H \equiv \psi^{\dagger} \hat{h} \psi$, c.f.\ Eq.~(\ref{H}), Eq.~(\ref{TRI}) implies that TRI imposes 
the condition $\sih{2} \kah{2} \ \hat{h}^{*} \, \sih{2} \kah{2} = \hat{h}$ upon the single-particle 
Hamiltonian $\hat{h}$.]\cite{footnote-b}
The product of the 
operations given by Eqs.~(\ref{PH}) and (\ref{TRI}) is a so-called ``chiral'' 
transformation,\cite{Zirnbauer,AltlandSimonsZirnbauer,BernardLeClair,GadeWegner,GuruswamyLeClairLudwig} 
and was denoted by the symbol ``$C_{z}$'' in Ref.~\onlinecite{OstrovskyGornyiMirlin1}; in this same 
reference, the physical TRI [Eq.~(\ref{TRI})] operation was labeled ``$T_{0}$.''

In addition, it is useful to define two additional transformations, which \emph{do not} correspond
to microscopic symmetry operations in the tight-binding honeycomb model of graphene, but nevertheless
may play important roles as \emph{emergent} symmetries at a special critical point, or when additional 
restrictions are placed upon the disorder distribution:
\begin{align}
	\psi(\vex{r}) &\rightarrow -\sih{1} \kah{2} 
	\left[
	\psi(\vex{r})^{\dagger}
	\right]^{\mathsf{T}},
	&&(\textrm{PH}^{*});\label{PHstar}
	\\
	\psi(\vex{r}) &\rightarrow \sih{2} \,
	\psi(\vex{r}),
	&&(\textrm{TRI}^{*}).\label{TRIstar}
\end{align}
The transformation $\textrm{PH}^{*}$ is defined to be unitary, while $\textrm{TRI}^{*}$ is antiunitary.
The product of $\textrm{PH}^{*}$ [Eqs.~(\ref{PHstar})] and the physical TRI operation [Eq.~(\ref{TRI})] is 
another type of ``chiral'' transformation, denoted by the symbol ``$C_{0}$'' in Ref.~\onlinecite{OstrovskyGornyiMirlin1}; 
the $\textrm{TRI}^{*}$ operation [Eq.~(\ref{TRIstar})] was labeled ``$T_{x}$'' in this reference.

The effective field theory for undoped graphene is encapsulated by the Hamiltonian
\begin{align}\label{H}
	H = &\int d^{2}\bm{\mathrm{r}} \,
	\psi^{\dagger}(\bm{\mathrm{r}})
	\left[
	-i v_{F} \sigmah \cdot \bm{\nabla}
	+ \hat{\mathcal{V}}(\bm{\mathrm{r}})
	\right]
	\psi(\bm{\mathrm{r}})
	\nonumber\\
	& + \frac{\mathcal{W}}{2}
	\int d^{2}\bm{\mathrm{r}} \, d^{2}\bm{\mathrm{r'}} \,
	\frac{
	\psi^{\dagger} \psi(\bm{\mathrm{r}}) \,
	\psi^{\dagger} \psi(\bm{\mathrm{r'}})
	}
	{|\bm{\mathrm{r}} - \bm{\mathrm{r'}}|}.
\end{align}
In this equation, $v_{F} \sim 3 t/2$ is the Fermi velocity, $\hat{\mathcal{V}}(\bm{\mathrm{r}})$ is a matrix-valued 
single-particle potential that encodes the structure of the quenched disorder (discussed below), and 
$\mathcal{W} \sim e^{2}/\epsilon$
is the microscopic strength of the Coulomb interactions.

We assume that both TRI [Eq.~(\ref{TRI})] and spin SU(2) rotational symmetry survive in every static 
realization of disorder. Our study thus excludes the effects of external magnetic fields or spin-flip
impurities, but is appropriate to remote charged impurities, 
``ripples,''\cite{MorozovNovoselov,MeyerGeim,MorpurgoGuinea,FasolinoLosKatsnelson} 
non-magnetic interstitial atoms, and
topological defects such as dislocations and bound disclination pairs. We will be interested in the properties
of a graphene system ensemble-averaged over many realizations of disorder; we presume that such averaging
restores invariance under PH transformations [Eq.~(\ref{PH})], as well as all honeycomb lattice space group 
symmetry operations (translations, rotations, reflections). Then the disorder potential 
$\hat{\mathcal{V}}(\bm{\mathrm{r}})$ appearing in Eq.~(\ref{H}) is conveniently parameterized as 
\begin{align}\label{DisDef}
	\mathcal{V}(\vex{r}) \equiv& 
	u(\vex{r})\,\hat{1}
	+A^{\betab}_{\alphab}(\vex{r}) \, \sih{\alphab} \kah{\betab}
	+A^{3}_{\alphab}(\vex{r}) \, \sih{\alphab} \kah{3}
	\nonumber\\
	&+ m^{\betab}(\vex{r})\,\sih{3}\kah{\betab} 
	+ v(\vex{r})\,\sih{3}\kah{3}.
\end{align}
In this equation, the barred indices $\alphab$ and $\betab$ are understood to be summed over the ``spatial''
Pauli matrix components
\begin{equation}\label{BarGIndDef}
	\alphab,\betab \in \{1,2\}.
\end{equation}
Note that Eq.~(\ref{DisDef}) implies that potentials $\{A^{\betab}_{\alphab}\}$ and $\{m^{\betab}\}$  
scatter between inequivalent Fermi nodes (i.e.\ couple to the valley space Pauli matrices $\kah{1}$ 
or $\kah{2}$); $u$, $\{A^{3}_{\alphab}\}$, and $v$ do not. Imposing statistical invariance under particle-hole 
[Eq.~(\ref{PH})]
and honeycomb lattice space group transformations, and assuming white noise (short-ranged), 
Gaussian-correlated disorder, we must set the average value of all disorder potentials in Eq.~(\ref{DisDef}) 
to zero, while we may assign up to five independent parameters $\{g_{u},g_{A},g_{A3},g_{m},g_{v}\}$ to 
characterize their statistical fluctuations:\cite{AleinerEfetov} 
\begin{subequations}\label{DisVar}
\begin{align}
	\overline{u(\vex{r})\,u(\vex{r'})} &= 
		2 \pi g_{u} \, v_{F}^2
		\, \delta^{(2)}(\vex{r}-\vex{r'}),\label{guDef}\\
	\overline{A^{\betab}_{\alphab}(\vex{r})\,A^{\gammab}_{\lambdab}(\vex{r'})} &= 
		2 \pi g_{A} \, v_{F}^2
		\, \delta_{\alphab,\lambdab} \delta^{\betab,\gammab} \delta^{(2)}(\vex{r}-\vex{r'}),\label{gADef}\\
	\overline{A^{3}_{\alphab}(\vex{r})\,A^{3}_{\lambdab}(\vex{r'})} &= 
		2 \pi g_{A3} \, v_{F}^2
		\, \delta_{\alphab,\lambdab} \delta^{(2)}(\vex{r}-\vex{r'}),\label{gA3Def}\\
	\overline{m^{\betab}(\vex{r})\,m^{\gammab}(\vex{r'})} &= 
		2 \pi g_{m} \, v_{F}^2
		\, \delta^{\betab,\gammab} \delta^{(2)}(\vex{r}-\vex{r'}),\label{gmDef}\\
	\overline{v(\vex{r})\,v(\vex{r'})} &= 
		2 \pi g_{v} \, v_{F}^2
		\, \delta^{(2)}(\vex{r}-\vex{r'}).\label{gvDef}
\end{align}
\end{subequations}

Let us now take a brief excursion from setting up our model to review the physical interpretation 
of these various disorder types described by Eqs.~(\ref{DisDef}) and (\ref{DisVar}); see also 
Refs.~\onlinecite{AleinerEfetov,OstrovskyGornyiMirlin1,TopologicalDisorder}. We will emphasize the connection between the particular 
types of disorder, the ``symmetry'' operations defined by Eqs.~(\ref{PH})--(\ref{TRIstar}), and the 
different universality classes\cite{Zirnbauer,AltlandSimonsZirnbauer,BernardLeClair} of disordered 
quantum systems realized by fine-tuning the disorder distribution in graphene.

\subsubsection{Scalar potential disorder and the symplectic class AII\label{ClassAII}}

Remote charged impurities at or near the substrate surface may provide the dominant scattering mechanism for electrons
in the experiments of Refs.~\onlinecite{Geim1,Geim2,Kim,MorozovNovoselov,Kim2}. 
One expects the potential arising from these impurities to vary relatively slowly on the graphene lattice scale, 
and therefore to manifest itself primarily through the effective scalar disorder potential $u(\vex{r})$ in the low-energy 
theory [Eqs.~(\ref{H})--(\ref{DisVar})]. We emphasize that the translation of any lattice-scale disorder potential 
into the single effective disorder field $u(\vex{r})$ is 
typically impossible, and all disorder types consistent 
with the underlying symmetry [i.e.\ time-reversal invariance, Eq.~(\ref{TRI})] will appear in the low-energy theory. 
Nevertheless, theoretically we are free to fine-tune the disorder distribution in Eqs.~(\ref{H}), (\ref{DisDef}), 
and (\ref{DisVar}) by setting
to zero all disorder parameters except $g_{u}$, i.e.\ only $u(\vex{r})$ nonzero. The resulting theory possesses 
a large SU(4) symmetry, present in every fixed disorder realization, associated with unitary transformations in 
$(\textrm{valley})\otimes(\textrm{spin-1/2})$ space. In every realization of $u(\vex{r})$, this theory is 
invariant under the alternative ``time-reversal'' transformation $\textrm{TRI}^{*}$, defined by 
Eq.~(\ref{TRIstar}). The advent of $\textrm{TRI}^{*}$ places the non-interacting system [Eqs.~(\ref{H})
and (\ref{DisDef}) with $\mathcal{W} = 0$ and only $u(\vex{r})\neq 0$] into the ``symplectic'' or 
spin-orbit universality class AII of disordered metals,\cite{LFSG,AleinerEfetov,OstrovskyGornyiMirlin1,
OstrovskyGornyiMirlin2,BardarsonTworzydloBrouwerBeenakker,NomuraKoshinoRyu,RyuMudryObuseFurusaki} rather than the 
orthogonal class AI nominally expected on the basis of spin-$1/2$ SU(2) rotational symmetry and ``physical'' TRI 
[Eq.~(\ref{TRI})].\cite{footnote-c}
(We have adopted the nomenclature for quantum disorder classes employed in Refs.~\onlinecite{Zirnbauer,AltlandSimonsZirnbauer,BernardLeClair}.)
Unlike the orthogonal class, the symplectic class is known to 
possess (in the absence of interactions) a disorder-driven metal-insulator transition in 2D.\cite{LeeRamakrishnan}
The (de-)localization of Dirac electrons subject only to scalar potential disorder has been studied in 
Refs.~\onlinecite{OstrovskyGornyiMirlin2,BardarsonTworzydloBrouwerBeenakker,NomuraKoshinoRyu,RyuMudryObuseFurusaki}.

\subsubsection{Vector potential disorder and ripples, topological defects, CI and BDI classes\label{ClassCIandBDI}}

In the context of graphene, the intravalley Abelian vector potential $\{A^{3}_{\alphab}\}$ in 
Eqs.~(\ref{H}) and (\ref{DisDef}) arises in the description of elastic lattice deformations or 
``ripples.''\cite{MorozovNovoselov,MeyerGeim,OstrovskyGornyiMirlin1,MorpurgoGuinea}
Ripples constitute a potentially important source of scattering in the electronic transport
of graphene, the presence of which has been directly observed and indirectly inferred in 
substrate-supported\cite{MorozovNovoselov} and suspended\cite{MeyerGeim} graphene samples, respectively. 
Topological lattice defects constitute another class of sample imperfections. In graphene, lattice 
dislocations can be modeled by point flux insertions of the potential $\{A^{3}_{\alphab}\}$ in the 
low-energy Dirac theory [Eq.~(\ref{H})], while the description of lattice disclinations requires the 
intervalley components $\{A^{\betab}_{\alphab}\}$.\cite{TopologicalDisorder}
Taken together, $\{A^{\betab}_{\alphab}\}$ and $\{A^{3}_{\alphab}\}$ form a quenched, non-Abelian 
SU(2) vector potential realized in valley space. 

The graphene field theory [Eqs.~(\ref{H}) and (\ref{DisDef})] possessing only quenched SU(2) vector potential 
disorder, obtained by fine-tuning $u(\vex{r}) = m^{\betab}(\vex{r}) = v(\vex{r}) =0$, is invariant (in every 
disorder realization) under the alternative ``particle-hole'' transformation $\textrm{PH}^{*}$, defined by 
Eq.~(\ref{PHstar}). In the absence of interparticle interactions [$\mathcal{W} = 0$ in Eq.~(\ref{H})], the 
corresponding disorder-averaged Dirac model with $g_{u} = g_{m} = g_{v} = 0$ in Eq.~(\ref{DisVar}) is a critical 
(i.e.\ not Anderson localized) conformal field theory (CFT), for which several exact and/or non-perturbative
results are known.\cite{NersesyanTsvelikWenger,CILogCFT} This CFT realizes a (variant of) the non-Wigner-Dyson 
universality class CI, typically associated with disordered superconductors.\cite{BernardLeClair,NersesyanTsvelikWenger,AltlandSimonsZirnbauer}
Intriguingly, the Dirac theory with only non-Abelian disorder is known to possess a vanishing density of states,\cite{NersesyanTsvelikWenger}
and is predicted to exhibit a nonzero conductance, independent of the disorder 
strength.\cite{OstrovskyGornyiMirlin1}
We note, however, that the $\textrm{PH}^{*}$ symmetry transformation given by Eq.~(\ref{PHstar}) does not 
correspond to any local operation in terms of the lattice electrons appearing in the model defined by Eq.~(\ref{H0}). 
As a consequence, neither elastic deformations nor topological defects occuring in the lattice-scale 
description of graphene translate into pure non-Abelian gauge disorder in the low-energy Dirac theory. 
For example, the field theory description of lattice dislocations and disclinations
incorporates 
the other disorder types $\{u(\vex{r}),m^{\betab}(\vex{r}),v(\vex{r})\}$ into the cores of the defects, 
so that an idealized graphene sheet subject only to topological disorder does \emph{not} realize the 
CI class CFT. 

Potentials $\{A_{\alphab}^{3}\}$ and $\{m^{\gammab}\}$ preserve both TRI and (physical)
particle-hole symmetry (PH) [Eq.~(\ref{PH})].
These particle-hole symmetric potentials \emph{can} be mapped
to a particular type of ``microscopic'' disorder consisting of real, nearest-neighbor random hopping
on the honeycomb lattice. Such disorder might be realized in principle in a graphene sheet
by introducing lattice vacancies, if the only effect of removing a carbon atom is to sever the 
3 nearest-neighbor electronic bonds surrounding the vacancy (i.e., treating the vacancy in the 
unitary limit). In the presence of TRI [Eq.~(\ref{TRI})] and spin SU(2) rotational symmetry, and
the absence of interactions ($\bar{w} = 0$) and further-neighbor hopping, such a model with only 
$g_{A3}$ and $g_{m}$ non-zero resides in the ``chiral'' symmetry class 
BDI,\cite{HatsugaiWenKohmoto,BernardLeClair,GuruswamyLeClairLudwig} 
and is another example of a critical, delocalized disordered-Dirac model in 2D for which exact and/or 
non-perturbative results are available.\cite{GuruswamyLeClairLudwig}

\subsubsection{Staggered on-site potential disorder and class D\label{ClassD}}

Finally, the disorder potential $v(\vex{r})$, which couples to a matrix diagonal in both sublattice ($\sigma$) and
valley ($\kappa$) spaces [Eq.~(\ref{DisDef})], may arise in concert with $u(\vex{r})$, e.g., in the 
description of interstitial impurities. The imposition of a \emph{uniform} $v(\vex{r}) \equiv v_{0}$ potential 
is consistent with the application of an on-site, $A$-$B$ sublattice-staggered chemical potential to the 
underlying honeycomb lattice model [Eq.~(\ref{H0})].
The non-interacting Dirac theory with only staggered potential disorder, Eqs.~(\ref{H}) and (\ref{DisVar}) 
with $\mathcal{W} = 0$ and only $g_{v} > 0$, realizes the non--Wigner-Dyson class D, and is formally related 
to the random bond Ising model in 2D.\cite{DotsenkoDotsenko,LFSG,SF,BSZ,RL}

We see that by fine-tuning the details of the disorder distribution within the confines of 
a graphene sheet that possesses, on average, the full space group symmetry of the honeycomb lattice,
several universality classes of (non-interacting) disordered quantum systems may be realized. The realization
of these classes is essentially equivalent to imposing various combinations of the 
discrete symmetries defined by Eqs.~(\ref{PH})--(\ref{TRIstar}). We emphasize, however,
that in the current situation of zero-magnetic field graphene experiments,\cite{Geim1,Geim2,MorozovNovoselov,Kim,Kim2} 
only the physical time-reversal symmetry [Eq.~(\ref{TRI})] presumably persists, and the ultimate low-energy, 
asymptotic physics must be described by the orthogonal Wigner-Dyson class AI.\cite{AleinerEfetov} Moreover, 
the Coulomb interactions are expected to play a strong role, as we will demonstrate in this paper.

We now complete the formal setup of our graphene field theory. We employ the zero-temperature,
imaginary time path integral formalism, implementing replicas $\psi \rightarrow \psi_{a}^{i}$
in order to average over the disorder. (The index $i \in \{1,2,\ldots,n\}$, with $n \rightarrow 0$
at the end of the calculation). The post-disorder averaged, imaginary time, coherent state path 
integral is
\begin{equation}\label{Zbar}
	\bar{Z} = \int \mathcal{D}\bar{\psi} \mathcal{D}\psi \, e^{-\bar{S}},
\end{equation}
with
\begin{align}\label{Sbar}
	\bar{S} =& \int d\tau \, d^{\textrm{2}}\bm{\mathrm{r}} \,
	\bar{\psi}^{i} 
	\left(\partial_{\tau} -i v_{F} \sigmah \cdot \bm{\nabla}\right)
	\psi^{i}
	\nonumber\\
	&+ 
	\frac{\mathcal{W}}{2}
	\int d\tau \, d^{2}\bm{\mathrm{r}} \, d^{2}\bm{\mathrm{r'}} \,
	\frac{
	\bar{\psi}^{i} \psi^{i}(\bm{\mathrm{r}}) \,
	\bar{\psi}^{i} \psi^{i}(\bm{\mathrm{r'}})
	}
	{|\bm{\mathrm{r}} - \bm{\mathrm{r'}}|}
	\nonumber\\
	&- 
	\frac{2 \pi g_{u} v_{F}^2}{2} \int d\tau \, d\tau' \, \dvex{r} \,
	\bar{\psi}^{i} \psi^{i}(\tau) \, \bar{\psi}^{j} \psi^{j}(\tau')
	\nonumber\\
	&-
	\frac{2 \pi \mathcal{G}^{\nu}_{\mu} v_{F}^2}{2} \int d\tau \, d\tau' \, \dvex{r} \,
	\bar{\psi}^{i} \sih{\mu} \kah{\nu} \psi^{i}(\tau) \, \bar{\psi}^{j} \sih{\mu} \kah{\nu} \psi^{j}(\tau'),
\end{align}	
where summation is implied over the repeated replica $\{i,j\}$ and disorder vertex $\{\mu,\nu\}$ indices 
(defined below). 
The disorder is encoded in the terms on the third and fourth lines of Eq.~(\ref{Sbar}), the
latter of which involves the disorder ``metric''
\begin{equation}\label{DisMetric}
	\mathcal{G}^{\nu}_{\mu} \rightarrow 
	\begin{bmatrix}
	g_{A} & g_{A} & g_{A3} \\
	g_{A} & g_{A} & g_{A3} \\
	g_{m} & g_{m} & g_{v}
	\end{bmatrix}
\end{equation}
[c.f.\ Eqs.~(\ref{DisDef}) and (\ref{DisVar})].
$\mathcal{G}^{\nu}_{\mu}$ couples to the direct product $\sih{\mu}\kah{\nu} \otimes \sih{\mu}\kah{\nu}$,
with $\mu,\nu \in \{1,2,3\}$.

Assuming that frequency carries the (generally scale-dependent) ``engineering'' dimension
\begin{equation}\label{dimA1}
	[\omega_{n}] \equiv z,
\end{equation}
measured in inverse-length units, we 
assign the dimension of the $\psi$ field as
\begin{equation}\label{dimA2}
	[\psi(\tau,\bm{\mathrm{r}})] \equiv 1.
\end{equation}
Then we find that the Fermi velocity $v_{F}$ and the Coulomb interaction strength $\mathcal{W}$
share the dimension
\begin{equation}
	[v_{F}] = [\mathcal{W}] = z - 1,\label{dimA3}
\end{equation}
while the disorder strengths $\{g_{u},\mathcal{G}^{\nu}_{\mu}\}$ are dimensionless.

\subsection{The theory at $N = \infty$\label{LargeN}}

The field theory given by Eqs.~(\ref{Zbar})--(\ref{DisMetric}) was studied by Stauber et al.\ in 
Ref.~\onlinecite{StauberGuineaVozmediano} for the case of weak disorder and weak Coulomb interactions. 
(Their results are discussed in Sec.~\ref{Results}.)
In graphene, the effective Coulomb interaction strength is of order unity, calling into question
the usefulness of a perturbative expansion in powers of $\mathcal{W}$ [Eq.~(\ref{Sbar})]. 
In this paper, we employ a large-$N$ generalization in order to treat the case of 
moderate to strong Coulomb interactions. The expansion parameter $1/N = 1/4$ for physical graphene,
which is at least smaller than the bare
dimensionless Coulomb interaction strength. 
Our treatment follows closely that of Ye in Ref.~\onlinecite{Ye},
wherein the author considered a large-$N$ generalization of a Dirac theory possessing a single 
valley. 

We generalize the Dirac spinor $\psi_{a}^{i}$ [Eq.~(\ref{PsiDef2})] to $2N \gg 1$ components in the index $a$, 
with physical graphene corresponding to $N = 4$.  [$i$ is the replica 
index, defined above Eq.~(\ref{Zbar}).] Crucially, we 
will consider only \emph{even} $N$, retaining the partioning of $N$ into two inequivalent Fermi point
sectors; as above, we will continue to use the Pauli matrices $\kah{m}$, $m\in\{1,2,3\}$, to address components 
in this valley space. In other words, we may write $N = 2(2s+1)$, where we take the spin $s \gg 1$---formally, 
this is a large spin expansion. At large $N$, we consider only the spin-independent, ``physical'' disorder 
potentials defined by Eqs.~(\ref{DisDef}) and (\ref{DisVar}), acting in the product of sublattice ($\sigma$) 
and valley ($\kappa$) spaces.

\begin{figure}
\includegraphics[width=0.4\textwidth]{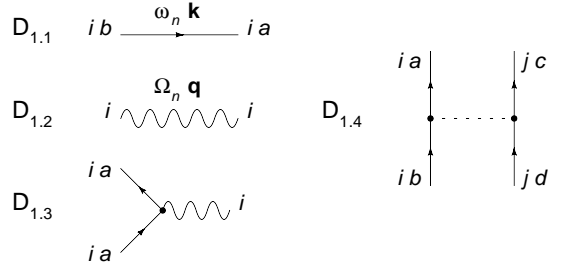}
\caption{Diagrammatic elements of the bare Feynman rules.
The associated amplitudes are summarized in Table~\ref{FR}.
\label{FigD3}}
\end{figure}

\begin{figure}[b]
\includegraphics[width=0.4\textwidth]{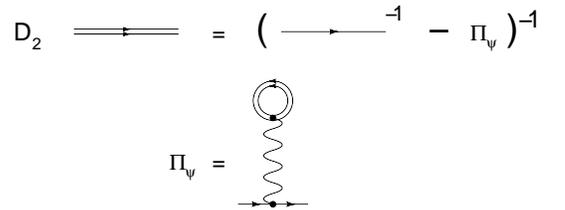}
\caption{The fermion propagator at $N = \infty$.
\label{FigD4}}
\end{figure}

We begin by decoupling the Coulomb interaction in Eqs.~(\ref{Zbar}) and (\ref{Sbar}) with a Hubbard-Stratonovich 
transformation, using the temporal gauge field $a^{i}$ ($i$ is again the replica index):
\begin{equation}\label{Zbar2}
	\bar{z} \rightarrow \int \mathcal{D}\bar{\psi} \mathcal{D}\psi \mathcal{D}a \, e^{-\bar{S}},
\end{equation}
where
\begin{align}\label{Sbar2}
	\bar{S} \rightarrow
	& \int d\tau \, d^{2}\bm{\mathrm{r}} \,
	\bar{\psi}^{i} 
	\left[\partial_{\tau} 
	-i \sqrt{\frac{2 w}{N}}a^{i}(\vex{r},0)
	-i v_{F} \sigmah \cdot \bm{\nabla}\right]
	\psi^{i}
	\nonumber\\
	&+ 
	\frac{1}{2}
	\int d\tau \, d^{2}\bm{\mathrm{r}} \, d z
	\left\lgroup
	\left[\nabla a^{i}(\vex{r},z)\right]^{2}
	+ 
	\left[\partial_{z} a^{i}(\vex{r},z)\right]^{2}
	\right\rgroup
	\nonumber\\
	&-
	\frac{2 \pi g_{u} v_{F}^2}{2} \int d\tau \, d\tau' \, \dvex{r} \,
	\bar{\psi}^{i} \psi^{i}(\tau) \, \bar{\psi}^{j} \psi^{j}(\tau')
	\nonumber\\
	&-
	\frac{2 \pi \mathcal{G}^{\nu}_{\mu} v_{F}^2}{2} \int d\tau \, d\tau' \, \dvex{r} \,
	\bar{\psi}^{i} \sih{\mu} \kah{\nu} \psi^{i}(\tau) \, \bar{\psi}^{j} \sih{\mu} \kah{\nu} \psi^{j}(\tau').
\end{align}
The temporal gauge field $a^{i}$ resides in the full $3+1$ spacetime $(\tau,\vex{r},z)$, while
the graphene field theory is confined to the plane $z = 0$. As a result, the coefficient 
in front of the kinetic term on the second line of Eq.~(\ref{Sbar2}) cannot be renormalized,
as we check explicitly in Sec.~\ref{photonSLFNRG}. 
In Eq.~(\ref{Sbar2}), we have replaced $\mathcal{W} \rightarrow w/2\pi N$; we keep
$w$ finite and non-zero so as to obtain a well-defined theory in the limit $N \rightarrow \infty$. 
The bare Feynman rules are stated in Fig.~\ref{FigD3} and in Table~\ref{FR}.
In this paper, we denote each diagram and its corresponding amplitude by the same symbol,
$\mathsf{D}_{m.n}$, where $m$ is the figure number and $n$ is the particular diagram in question.

\begin{table}[b]
\caption{
	\label{FR} 
	Factors associated to bare propagators and vertices depicted in Fig.~\ref{FigD3}.
	}
\begin{ruledtabular}
\begin{tabular*}{\textwidth}{@{\extracolsep{\fill}}lll}
\smallskip\\
$\mathsf{D}_{1.1}$ & $=$ 
&
$\displaystyle{
\frac{
	(
		i \omega_{n} + v_{F} \sigmah \cdot \bm{\mathrm{k}}
	)_{a,b}
	}
	{\omega_{n}^{2} + v_{F} k^{2}}
}$
\medskip\\
$\mathsf{D}_{1.2}$ & $=$ 
&
$\displaystyle{
\frac{1}{q}
}$
\medskip\\
$\mathsf{D}_{1.3}$ & $=$ 
&
$\displaystyle{
i \sqrt{\frac{w}{N}}
}$
\medskip\\
$\mathsf{D}_{1.4}$ & $=$ 
&
$\displaystyle{
\begin{array}{l}
2 \pi g_{u} v_{F}^2 \delta_{a,b} \delta_{c,d} \\
+ 2 \pi \mathcal{G}^{\nu}_{\mu} v_{F}^2 (\sih{\mu}\kah{\nu})_{a,b}(\sih{\mu}\kah{\nu})_{c,d}
\end{array}
}$
\bigskip\\
\end{tabular*}
\end{ruledtabular}
\end{table}

\begin{figure}[t]
\includegraphics[width=0.4\textwidth]{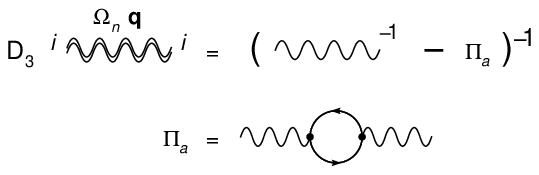}
\caption{The gauge (Coulomb) propagator at $N = \infty$.
\label{FigD5}}
\end{figure}

The $N \rightarrow \infty$ theory, in the absence of disorder, is completely
characterized by the one- and two-particle self-energies depicted in Figs.~\ref{FigD4} and
\ref{FigD5}, respectively. For the special case of massless Dirac fermions studied here,
however, particle-hole symmetry prevents the generation of a chemical potential shift, so that
the electronic self-energy $\Pi_{\psi}$ vanishes at $N = \infty$ (and zero disorder), 
and the large-$N$ fermion propagator is the same as $\mathsf{D}_{1.1}$ (Fig.~\ref{FigD3} and Table~\ref{FR}).

The gauge field self-energy at $N = \infty$ is depicted in Fig.~\ref{FigD5}; it is given by the expression
\begin{equation}\label{PiaNinf}
	\Pi_{a}(\Omega_{n},\bm{\mathrm{q}}) = (i \sqrt{w/N})^{2} \mathcal{D}_{0}(\Omega_{n},\bm{\mathrm{q}}),
\end{equation}
where the Dirac fermion polarization bubble $\mathcal{D}_{0}(\Omega_{n},\bm{\mathrm{q}})$ is
\begin{align}\label{D0psi}
	\mathcal{D}_{0}(\Omega_{n},\bm{\mathrm{q}}) & = 
	\frac{N}{16}
	\frac{q^{2}}{\sqrt{\Omega_{n}^{2} + v_{F}^2 q^{2}}}.
\end{align}

Incorporating the self-energy from Eq.~(\ref{PiaNinf}), the propagator for the 
temporal gauge field at $N = \infty$ and zero disorder is 
\begin{equation}\label{GaugePropNinf}
	\mathsf{D}_{3} = 
	\frac{1}{q} \frac{\sqrt{v_{F}^2 q^{2} + \Omega_{n}^{2}}}
	{\sqrt{v_{F}^2 q^{2} + \Omega_{n}^{2}} + \bar{w} v_{F} |\vex{q}|},
\end{equation}
where we have introduced the dimensionless Coulomb interaction strength
\begin{equation}\label{wbar}
	\bar{w} \equiv \frac{w}{16 v_{F}}
\end{equation}
[c.f.\ Eq.~(\ref{dimA3})]. 
For the physical case of $N = 4$, the dimensionless parameter $\bar{w}$ is essentially
the Coulomb interaction constant $r_{s}$:
\begin{equation}\label{wbarrs}
	\bar{w} = \frac{\pi}{2} \frac{e^{2}}{\epsilon v_{F}}.
\end{equation}

\begin{figure}[b]
\includegraphics[width=0.4\textwidth]{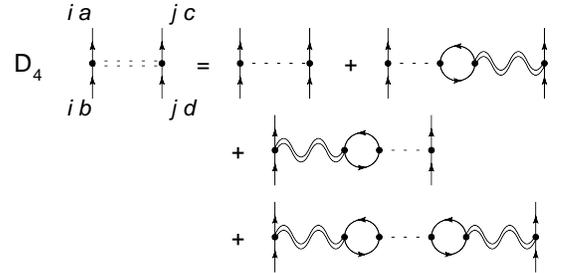}
\caption{Screening of the scalar potential disorder $g_{u}$ due to 
Coulomb interactions in the $N = \infty$ limit.
\label{FigD6}}
\end{figure}

The scalar potential disorder, characterized by the disorder strength $g_{u}$ 
[Eqs.~(\ref{DisDef}) and (\ref{DisVar}), Fig.~\ref{FigD3} and Table~\ref{FR}]
is 
parametrically screened by the Dirac sea in the large $N$ limit; the screening
affects only the scalar potential disorder, since the other disorder fields
in Eq.~(\ref{DisVar}) can be understood as arising from charge-neutral defects 
that produce potentials lacking a monopole (i.e.\ lowest-order multipole) moment.
The diagram $\mathsf{D}_{4}$ pictured in Fig.~\ref{FigD6} gives the complete
dressed disorder vertex at $N = \infty$:
\begin{align}\label{DisVertexNinf}
	\mathsf{D}_{4}  
	\equiv& 2 \pi \, \tgu v_{F}^2\, \delta_{a,b} \delta_{c,d} 
	\nonumber\\
	&+ 2 \pi \, \mathcal{G}^{\nu}_{\mu} v_{F}^2\, (\sih{\mu}\kah{\nu})_{a,b}(\sih{\mu}\kah{\nu})_{c,d}, 
\end{align}
where we have defined the screened scalar potential disorder strength
\begin{equation}\label{tguDef}
	\tgu \equiv \frac{g_{u}}{(1 + \bar{w})^{2}}.
\end{equation}
In the large-$N$ theory, it is the screened strength $\tgu$, rather than the bare
parameter $g_{u}$, that appears in physical observables (Sec.~\ref{Physics} and
Appendix~\ref{APP-Physics}), and 
therefore characterizes the effective strength of scalar potential fluctuations.

In summary, the Feynman rules for the large-$N$ version of graphene are given by 
$\mathsf{D}_{1.1}$ and $\mathsf{D}_{1.3}$ (Fig.~\ref{FigD3} and Table~\ref{FR}),
and by $\mathsf{D}_{3}$ and $\mathsf{D}_{4}$ [Figs.~\ref{FigD5} and \ref{FigD6},
Eqs.~(\ref{GaugePropNinf}) and (\ref{DisVertexNinf}), respectively].


\section{Perturbative expansion\label{PerturbativeCalc}}

In this section, we perform a one-loop renormalization group (RG) calculation within the large-$N$
generalization of graphene defined by Eqs.~(\ref{Zbar2}) and (\ref{Sbar2}). Using the results
derived below, we obtain flow equations for the disorder and interaction coupling strengths defined 
by Eqs.~(\ref{H})--(\ref{DisVar}). These flow equations are stated and analyzed in Sec.~\ref{Results}.

\subsection{Electronic self-energy}\label{SLFNRG}

\begin{figure}
\includegraphics[width=0.3\textwidth]{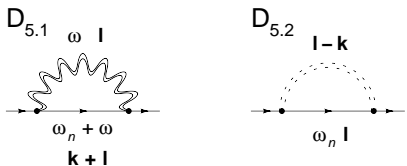}
\caption{Corrections to the electronic self-energy at $\ord{\frac{1}{N},\{\tgu,\mathcal{G}^{\nu}_{\mu}\}}$.
\label{FigD7}}
\end{figure}

We begin with the renormalization of the electronic self-energy due to the disorder
and Coulomb interactions (at finite, but large $N$). The diagrams are depicted
in Fig.~\ref{FigD7}. We consider first the $1/N$ correction due to the interaction,
diagram $\mathsf{D}_{5.1}$. Expanding in powers of the external frequency $\omega_{n}$ 
and momentum $\vex{k}$, we find that 
\begin{align}\label{D7.1}
	\mathsf{D}_{5.1}
	&\sim 
	- \eta
	\left[\bar{w} f_{1}(\bar{w}) i \omega_{n} +\bar{w} f_{2}(\bar{w}) v_{F} \sigmah\cdot\vex{k}\right]
	\ln\Lambda,
\end{align}
where $\bar{w}$ is the dimensionless Coulomb interaction strength defined by Eq.~(\ref{wbar}),
$\Lambda$ is a hard momentum cutoff, and we have introduced 
\begin{equation}\label{etaDef}
	\eta \equiv \frac{8}{\pi N}.
\end{equation}
In Eq.~(\ref{D7.1}), the functions $f_{1}(\bar{w})$ and $f_{2}(\bar{w})$ denote the following 
analytic functions of the interaction strength $\bar{w}$
\begin{subequations}\label{f1f2Def}
\begin{align}
	f_{1}(\bar{w}) =& 
	\frac{1}{2}\frac{1}{\sqrt{1 - \bar{w}^{2}}}
	\left(
	\frac{1-\sqrt{1 - \bar{w}^{2}}}{1+\sqrt{1 - \bar{w}^{2}}}
	\right)
	\nonumber\\
	&-\frac{2}{\pi \bar{w}^{2}}
	\left[
	\frac{1 - \frac{\bar{w}^{2}}{2}}{\sqrt{1-\bar{w}^{2}}}\arcsin(\bar{w}) - \bar{w}
	\right],
	\label{f1Def}\\
	f_{2}(\bar{w}) =& 
	\frac{1}{2}\frac{1}{1+\sqrt{1 - \bar{w}^{2}}}
	\nonumber\\
	&+\frac{1}{\pi \bar{w}^{2}}
	\left[
	\sqrt{1-\bar{w}^{2}}\arcsin(\bar{w}) - \bar{w}
	\right].
	\label{f2Def}
\end{align}
\end{subequations}
It will prove useful to consider the sum
\begin{align}\label{f3Def}
	f_{3}(\bar{w}) \equiv&
	f_{1}(\bar{w}) + f_{2}(\bar{w})
	\nonumber\\
	=&
	\frac{1}{2\bw^{2}}
	\left\lgroup
	\frac{2}{\pi}
	\left[
	\bar{w} + \frac{\arccos(\bar{w})}{\sqrt{1-\bar{w}^{2}}}
	\right]
	 - 1 
	\right\rgroup.
\end{align}
The electronic self-energy contribution detailed in Eqs.~(\ref{D7.1})--(\ref{f3Def}) has
been considered several times before in the literature, first by Ye and Sachdev,\cite{YeSachdev,Ye} 
and soon after by Gonz\'alez, Guinea, and Vozmediano.\cite{GGV} 
The interaction self-energy at order $1/N$ was revisited recently by Son.\cite{Son}
Our results agree 
with those of Refs.~\onlinecite{GGV,Son}, but disagree with those of Refs.~\onlinecite{YeSachdev,Ye}. 
We derive Eqs.~(\ref{D7.1})--(\ref{f3Def}) explicitly in Appendix~\ref{APP-PsiSLFNRG}.

Next, we turn to the disorder diagram $\mathsf{D}_{5.2}$. We find 
\begin{align}\label{D7.2}
	\mathsf{D}_{5.2} 
	&\sim \bGs i \omega_{n} \ln\Lambda,
\end{align}	
where
\begin{align}\label{bGsDef}
	\bGs \equiv& \tgu + \sum_{\mu,\nu} \mathcal{G}^{\nu}_{\mu}
	\nonumber\\
	=& \tgu + 4 g_{A} + 2 g_{A3} + 2 g_{m} + g_{v}.
\end{align}
[$\tgu$ is the screened scalar potential disorder strength,
defined by Eq.~(\ref{tguDef})].

The bare irreducible two-point vertex function is
\begin{align}
	i \Gamma_{\psi\psi}^{(0)} &= i \omega_{n} - v_{F} \sigmah\cdot\vex{k} + \mathsf{D}_{5.1} + \mathsf{D}_{5.2},
\end{align}
and the renormalization condition is\cite{DJA}
\begin{equation}\label{PsiSLFNRGRenormCond}
	\dlambda \left( Z_{\psi}^{\frac{2}{2}} i \Gamma_{\psi\psi}^{(0)} \right)= 0,
\end{equation}
where $Z_{\psi}$ is the wavefunction renormalization of the $\psi$ field.
Using Eqs.~(\ref{D7.1}) and (\ref{D7.2}) in Eq.~(\ref{PsiSLFNRGRenormCond}),
and combining the result with the dimensional analysis [Eqs.~(\ref{dimA1})--(\ref{dimA3})],
we obtain the flow equations
\begin{align}
	\frac{d \ln Z_{\psi}}{d l} & =
	-\eta \bar{w} f_{1}(\bar{w}) + \bGs,
	\label{ZPsiRG}\\
	\frac{d \ln v_{F}}{d l} & =
	z - 1
	- \bGs
	+ \eta \bar{w} f_{3}(\bar{w}),
	\label{lnvFRG}
\end{align}
where $l = -\ln\Lambda$ is the log of the RG length scale,
and $z$ is the (as yet unspecified) ``dynamic critical exponent''
[Eq.~(\ref{dimA1})]. 

Although we are employing a field-theoretic renormalization scheme, 
Eqs.~(\ref{ZPsiRG}) and (\ref{lnvFRG}) can be equivalently understood
in a Wilsonian framework: after an integration of high-momentum modes,
one acquires corrections 
akin to those expressed in Eqs.~(\ref{D7.1}) and (\ref{D7.2}). 
[Consult the action given by Eq.~(\ref{Sbar}).] Rescaling momentum $k\rightarrow k/b$, 
frequency $\omega_{n}\rightarrow \omega_{n}/b^{z}$, and 
field  $\psi \rightarrow Z_{\psi}^{-1/2} \psi$, one recovers Eqs.~(\ref{ZPsiRG}) and 
(\ref{lnvFRG}), where $b \sim 1 + dl$.

\subsection{Coulomb vertex}\label{CoulombVertex}

\begin{figure}
\includegraphics[width=0.4\textwidth]{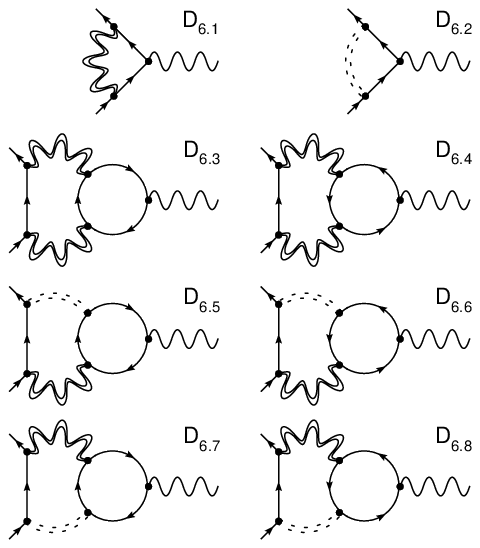}
\caption{Coulomb vertex corrections at $\ord{\frac{1}{N},\{\tgu,\mathcal{G}^{\nu}_{\mu}\}}$
\label{FigD8}}
\end{figure}

\begin{figure}[b]
\includegraphics[width=0.4\textwidth]{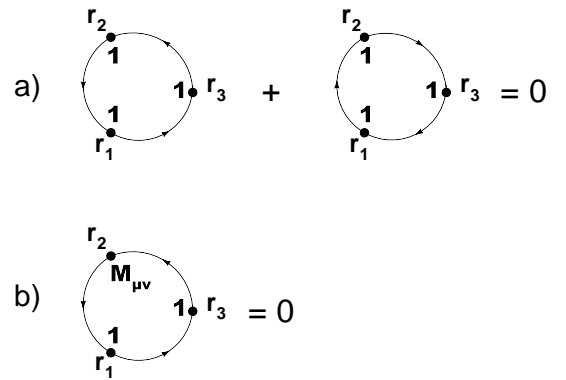}
\caption{a) Cancelation between counter-propagating 3-electron rings involving only identity matrix ($\bm{1}$)
insertions into the trace. b) 3-electron rings involving two identity matrix insertions and one disorder matrix
of the form $\hat{{\mathrm{M}}}_{\mu \nu} = \sih{\mu} \kah{\nu}$ vanish identically.
\label{FigD9}}
\end{figure}

One-loop corrections to the Coulomb vertex $\mathsf{D}_{1.3}$ (Fig.~\ref{FigD3} and
Table~\ref{FR}) are depicted in Fig.~\ref{FigD8}. Most of these diagrams involve a ``3-electron ring,'' 
that is, a closed electron loop with three nodes. 
Such structures are generic to large-$N$ expansions.
We note that the graphene (massless Dirac) theory studied here possesses two simplifying features: 
\begin{itemize}
\item[a)]{
	The sum of two counter propagating, but otherwise identical, $m$-electron rings
	\emph{vanishes} for odd $m$ \emph{whenever all $m$ bilinear insertions 
	involve the identity matrix} $\hat{1}$; see Fig.~\ref{FigD9}a. This result is 
	simply a consequence of the odd parity nature of the massless Dirac Green's function
	$\hat{G}(\vex{r})$:
	\begin{equation}\label{3RingCancel}
		\hat{G}(\vex{r} - \vex{r'}) = - \hat{G}(\vex{r'} - \vex{r}).
	\end{equation}
}
\item[b)]{
	An $m$-electron ring involving $m-1$ identity matrix insertions and 
	a single disorder matrix insertion of the form 
	$\hat{\mathrm{M}}_{\mu \nu} \equiv \sih{\mu}\kah{\nu}$
	vanishes trivially by the trace over valley ($\kappa$) space; see 
	Fig.~\ref{FigD9}b.
}
\end{itemize}

As a result of this simplification, only diagrams $\mathsf{D}_{6.1}$ and $\mathsf{D}_{6.2}$
contribute. We find that
\begin{equation}\label{D8.1}
	\mathsf{D}_{6.1} \sim -\left(i \sqrt{\frac{w}{N}}\right)\eta \bar{w} f_{1}(\bar{w}) \ln\Lambda,
\end{equation}
and
\begin{align}\label{D8.2}
	\mathsf{D}_{6.2} 
	&\sim
	\left(i \sqrt{\frac{w}{N}}\right)
	\bGs \ln\Lambda.
\end{align}
In these equations, $f_{1}(\bar{w})$ was defined by Eq.~(\ref{f1Def}), 
$\eta$ by Eq.~(\ref{etaDef}), and $\bGs$ by Eq.~(\ref{bGsDef}).
The evaluation of Eq.~(\ref{D8.1}) closely parallels that of the electronic self-energy,
the calculation of which is detailed in Appendix~\ref{APP-PsiSLFNRG}.

The bare vertex function is 
\begin{equation}
	i \Gamma_{a\psi\psi}^{(0)} \sim 
	\left(i \sqrt{\frac{w}{N}}\right)
	\left\lgroup
	1 + \ln\Lambda 
	\left[
	\bGs-\eta \bar{w} f_{1}(\bar{\omega})
	\right]	
	\right\rgroup.
\end{equation}
Now, we need to know the wavefunction renormalizations for both the Dirac ($\psi$)
and gauge ($a$) fields to proceed; the latter requires the evaluation of the 
gauge field self-energy to one loop. 
Gauge invariance in fact requires that 
\begin{equation}\label{ZaRG}
	\frac{d \ln Z_{a}}{d \ln\Lambda} = 0.
\end{equation}
We will 
check Eq.~(\ref{ZaRG}) in Sec.~\ref{photonSLFNRG}, below.

The renormalization condition is therefore
\begin{equation}
	\left[\dlambda + \frac{d \ln Z_{\psi}}{d \ln\Lambda}\right] i \Gamma_{a\psi\psi}^{(0)} = 0,
\end{equation}
which implies the following RG flow equation for the 
dimensionless Coulomb interaction strength $\bar{w}$ [introduced
in Eq.~(\ref{wbar})]:
\begin{align}\label{wbarRG}
	\frac{d \ln \bar{w}}{d l} 
	&=
	\bGs - \eta \bar{w} f_{3}(\bar{w})
\end{align}
[$l = -\ln\Lambda$ in this equation].
To obtain Eq.~(\ref{wbarRG}), we have used Eqs.~(\ref{f3Def}),
(\ref{ZPsiRG}), and (\ref{lnvFRG}).

\subsection{Gauge field self-energy}\label{photonSLFNRG}

\begin{figure}
\includegraphics[width=0.4\textwidth]{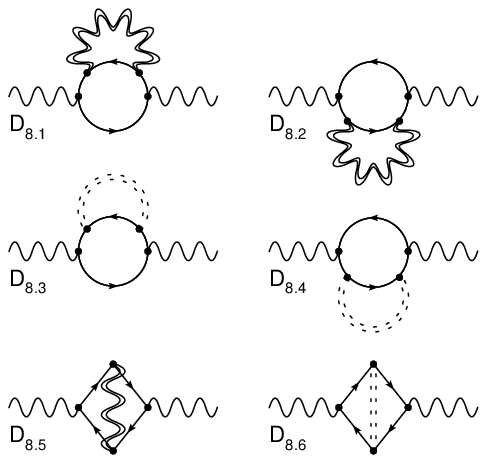}
\caption{Corrections to the gauge (Coulomb) field self-energy to 
$\ord{\frac{1}{N},\{\tgu,\mathcal{G}^{\nu}_{\mu}\}}$.
\label{FigD10}}
\end{figure}

\begin{figure}[b]
\includegraphics[width=0.48\textwidth]{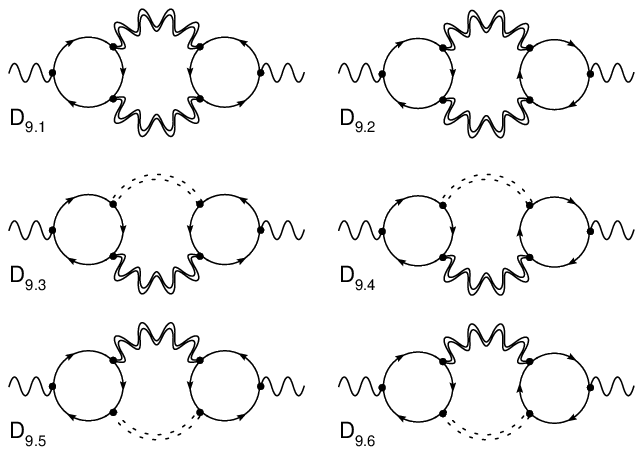}
\caption{Corrections to the gauge (Coulomb) field self-energy to 
$\ord{\frac{1}{N},\{\tgu,\mathcal{G}^{\nu}_{\mu}\}}$.
\label{FigD11}}
\end{figure}

\begin{figure}
\includegraphics[width=0.48\textwidth]{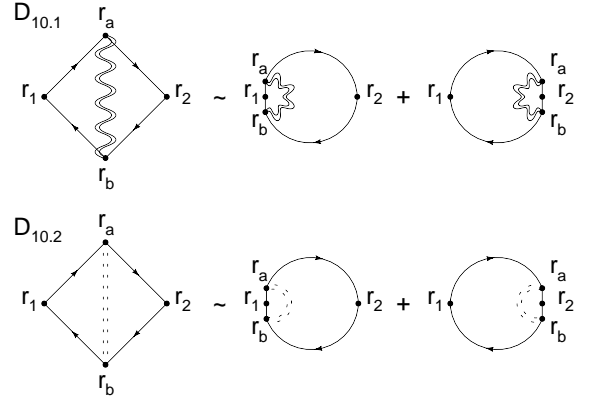}
\caption{Evaluation of ``two-loop'' diagrams $\mathsf{D}_{8.5}$ and $\mathsf{D}_{8.6}$ in Fig.~\ref{FigD10}.
These graphs can be expressed in terms of products of an ultraviolet-convergent polarization bubble and 
an ultraviolet-divergent vertex correction (see Fig.~\ref{FigD8}). 
\label{FigD12}}
\end{figure}

The diagrammatic corrections to the gauge field self-energy to are depicted in 
Figs.~\ref{FigD10} and \ref{FigD11}. Because of the properties explicated in Fig.~\ref{FigD9}, 
the sum of the diagrams in Fig.~\ref{FigD11} gives zero. 

All of the diagrams in Fig.~\ref{FigD10} are formally two-loop, but nevertheless
give only contributions to first order in $\ln\Lambda$. In other words, one
of the two loops in each of these diagrams is always ultraviolet-finite, and
we can obtain their contributions to the renormalization group using the
hard momentum cutoff scheme implemented above.
 
We begin with $\mathsf{D}_{8.1}$ and $\mathsf{D}_{8.2}$, which give
identical contributions. One finds that
\begin{align}\label{D10.1n2}
	\mathsf{D}_{8.1} + \mathsf{D}_{8.2} 
	\sim&
	2 A \bar{w} \frac{v_{F} q^{2}}{\sqrt{\Omega_{n}^{2}+ v_{F}^2 q^{2}}}
	+2 B \bar{w} \frac{16}{N} v_{F} J(\Omega_{n},\vex{q}),
\end{align}
where we have used Eq.~(\ref{D0psi}), and the log-divergent constants 
$A$ and $B$ follow from Eq.~(\ref{D7.1}):
\begin{subequations}\label{AB}
\begin{align}
	A &=
	\frac{1}{2}\eta\bar{w}[f_{2}(\bar{w})-f_{1}(\bar{w})]\ln\Lambda,
\end{align}
\begin{align}
	B &=
	-\frac{1}{2}\eta\bar{w} f_{3}(\bar{w}) \ln\Lambda.
\end{align}
\end{subequations}
In Eq.~(\ref{D10.1n2}), we have defined
\begin{align}\label{JDef}	
	&J(\Omega_{n},\vex{q})
	\nonumber\\
	&\quad\equiv \int 
	\frac{d\omega\,\dvex{l}}{(2\pi)^{3}}\,
	\frac
	{\Tr\left\lgroup
	\begin{aligned}
	&[i (\omega+\Omega_{n}) + v_{F} \sigmah\cdot(\vex{l}+\vex{q})]
	\\
	&\times[i \omega + v_{F} \sigmah\cdot\vex{l}]^{3}
	\end{aligned}
	\right\rgroup}
	{\left[\omega^{2}+v_{F}^2 \vex{l}^{2}\right]^{2}
	\left[(\omega+\Omega_{n})^{2}+v_{F}^2 (\vex{l}+\vex{q})^{2}\right]}
	\nonumber\\
	&\quad= 
	\frac{N}{16}
	\frac{q^{2} \Omega_{n}^{2}}{(v_{F}^2 q^{2}+\Omega_{n}^{2})^{\frac{3}{2}}}.
\end{align}

Next, we evaluate 
\begin{align}\label{D10.3n4}
	\mathsf{D}_{8.3} + &\mathsf{D}_{8.4}
	\nonumber\\
	&=
	2 A' \bar{w} \frac{v_{F} q^{2}}{\sqrt{\Omega_{n}^{2}+v_{F}^2 q^{2}}}
	+2 B' \bar{w} \frac{v_{F} q^{2} \Omega_{n}^{2}}{\left(\Omega_{n}^{2}+v_{F}^2 q^{2}\right)^{\frac{3}{2}}}.
\end{align}
In Eq.~(\ref{D10.3n4}), the log-divergent constants $A'$ and $B'$ 
follow from Eq.~(\ref{D7.2}):
\begin{align}\label{ABPRIME}
	A' = B'
	&=\frac{1}{2}\bGs \ln\Lambda.
\end{align}

Diagrams $\mathsf{D}_{8.5}$ and $\mathsf{D}_{8.6}$ can be evaluated 
by simply multiplying the divergent vertex correction [Eqs.~(\ref{D8.1}) and (\ref{D8.2})] 
(evaluated at the left \emph{and} right sides of the two-loop graphs, as shown in Fig.~\ref{FigD12}) by
a factor of the non-divergent polarization bubble, given by the $N = \infty$ photon
self-energy [Eq.~(\ref{PiaNinf})]. 
We obtain
\begin{subequations}\label{D10.5n6}
\begin{align}
	\mathsf{D}_{8.5} &\sim 2 \eta \bar{w}^{2} f_{1}(\bar{w}) \ln\Lambda\frac{v_{F} q^{2}}{\sqrt{\Omega_{n}^{2}+v_{F}^2 q^{2}}},
	\\
	\mathsf{D}_{8.6} &\sim -2 \bar{w} \bGs \ln\Lambda\frac{v_{F} q^{2}}{\sqrt{\Omega_{n}^{2}+v_{F}^2 q^{2}}}.
\end{align}
\end{subequations}

Summing the results of Eqs.~(\ref{D10.1n2}), (\ref{D10.3n4}), and (\ref{D10.5n6}) with the inverse
$N = \infty$ gauge propagator [Eq.~(\ref{GaugePropNinf})], and using Eqs.~(\ref{AB}) and (\ref{ABPRIME}), 
we obtain the bare vertex function
\begin{align}
	i \Gamma_{a a}^{(0)} =& 
	-|\vex{q}| - \bar{w} \frac{v_{F} q^{2}}{\sqrt{\Omega_{n}^{2}+v_{F}^2 q^{2}}}
	\nonumber\\
	&+ \frac{v_{F} q^{2}}{\sqrt{\Omega_{n}^{2}+v_{F}^2 q^{2}}}
	\left[
	\eta \bar{w}^{2} f_{3}(\bar{w}) - \bar{w} \bGs
	\right] \ln\Lambda
	\nonumber\\
	&+ 
	\frac{v_{F} q^{2} \Omega_{n}^{2}}{\left(\Omega_{n}^{2} + v_{F}^2 q^{2}\right)^{\frac{3}{2}}}
	\left[
	-\eta \bar{w}^{2} f_{3}(\bar{w}) + \bar{w} \bGs
	\right] \ln\Lambda.
\end{align}
Since there is no correction proportional to $|\vex{q}|$, we see immediately that
\begin{equation}\label{ZaRG--II}
	\frac{d \ln Z_{a}}{d \ln\Lambda} = 0,
\end{equation}
as assumed above in Eq.~(\ref{ZaRG}).
The renormalization condition is then
\begin{equation}\label{lnhandbarwRGRepeat}
	\dlambda i \Gamma_{a a}^{(0)} = 0.
\end{equation}
Eq.~(\ref{lnhandbarwRGRepeat}) recovers the previously-derived results,
Eqs.~(\ref{lnvFRG}) and (\ref{wbarRG}).

\subsection{Disorder strength renormalization}\label{DisRenormSec}

\begin{figure}[t]
\includegraphics[width=0.35\textwidth]{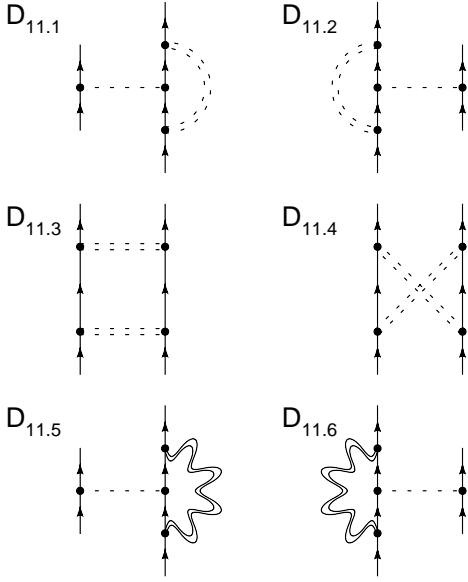}
\caption{Renormalization of the disorder strength at 
$\ord{\frac{g}{N},g^{2}}$, where $g \in \{\tgu,\mathcal{G}^{\nu}_{\mu}\}$.
\label{FigD13}}
\end{figure}

\begin{figure}
\includegraphics[width=0.4\textwidth]{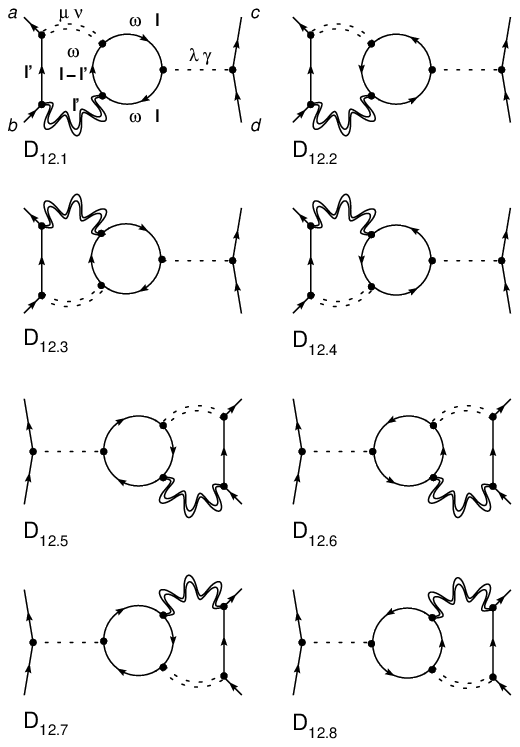}
\caption{Renormalization of the disorder strength at 
$\ord{\frac{g}{N},g^{2}}$, where $g \in \{\tgu,\mathcal{G}^{\nu}_{\mu}\}$.
\label{FigD14}}
\end{figure}

\begin{figure}[b]
\includegraphics[width=0.4\textwidth]{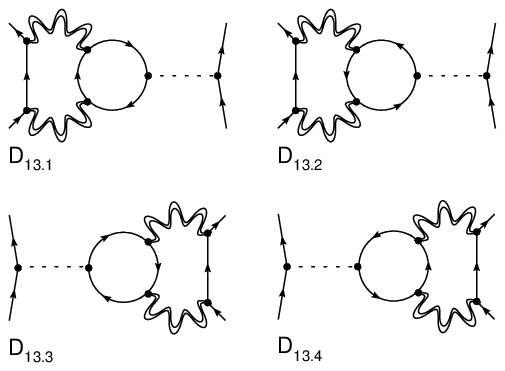}
\caption{Renormalization of the disorder strength at 
$\ord{\frac{g}{N},g^{2}}$, where $g \in \{\tgu,\mathcal{G}^{\nu}_{\mu}\}$.
\label{FigD15}}
\end{figure}

Finally, we compute the renormalization of the disorder strength, using
the diagrams pictured in Fig.~\ref{FigD13}--\ref{FigD15}. These diagrams
correct the \emph{bare} disorder vertex $\mathsf{D}_{1.4}$,
pictured in Fig.~\ref{FigD3}, with the tree-level amplitude given in Table~\ref{FR}.

Diagrams $\mathsf{D}_{11.1}$--$\mathsf{D}_{11.4}$ describe the autorenormalization of the disorder
at one loop. One finds\cite{AleinerEfetov}
\begin{align}\label{D13.1to4}
	\mathsf{D}_{11.1}&+\mathsf{D}_{11.2}+\mathsf{D}_{11.3}+\mathsf{D}_{11.4}
	\nonumber\\
	=&
	2 \pi v_{F}^2 \left(\hat{1}\otimes\hat{1}\right) 
	\left[
		2 g_{u} \bGs
		+ 8 g_{m} g_{A}
		+ 4 g_{v} g_{A3}  
	\right]\ln \Lambda
	\nonumber\\
	&+ 2 \pi v_{F}^2 \left(\sih{\alphab}\kah{\betab}\otimes\sih{\alphab}\kah{\betab}\right)
	\left[
	\begin{aligned}
	&8 g_{A3} g_{A} 
	\\
	&+ 2 g_{m} (\tgu+g_{v})
	\end{aligned}
	\right]\ln \Lambda
	\nonumber\\
	&+ 2 \pi v_{F}^2 \left(\sih{\alphab}\kah{3}\otimes\sih{\alphab}\kah{3}\right)
	\left[2 g_{m}^{2} + 8 g_{A}^{2} + 2\tgu g_{v}\right]\ln \Lambda
	\nonumber\\
	&+ 2 \pi v_{F}^2 \left(\sih{3}\kah{\betab}\otimes\sih{3}\kah{\betab}\right) 
	\left[
	\begin{aligned}
	&2g_{m}(g_{v}-\tgu) 
	\\
	&+ 4 g_{A} (\tgu + g_{v}) 
	\end{aligned}
	\right]\ln \Lambda
	\nonumber\\
	&+ 2 \pi v_{F}^2 \left(\sih{3}\kah{3}\otimes\sih{3}\kah{3}\right)
	\left[
	\begin{aligned}
	&2 g_{v}(2 g_{m} - \tgu - g_{v}) 
	\\
	&+ 4 g_{A3} (\tgu + g_{v}) 
	\\
	&+ 8 g_{A} (g_{m} - g_{v}) 
	\end{aligned}
	\right]\ln \Lambda,
\end{align}
where we have re-introduced the following notation:
barred Greek indices run over the ``spatial'' Pauli matrix components
\begin{equation}\label{BarGIndDef2}
	\alphab,\betab \in \{1,2\},
\end{equation}
to be distinguished from unbarred Greek indices which run over all
three Pauli matrix components, e.g.\
\begin{equation}\label{UnBarGIndDef}
	\alpha,\beta \in \{1,2,3\}.
\end{equation}

Diagrams $\mathsf{D}_{11.5}$ and $\mathsf{D}_{11.6}$ give identical contributions,
involving the dressing of the disorder vertex by the Coulomb interactions at 
$\ord{1/N}$. Their evaluation proceeds similarly to that of the electronic self-energy 
(detailed in Appendix~\ref{APP-PsiSLFNRG}). We obtain
\begin{align}\label{D13.5and6}
	\mathsf{D}_{11.5}&+\mathsf{D}_{11.6}
	\nonumber\\
	=&
	2 \pi g_{u} v_{F}^2 \left(\hat{1}\otimes\hat{1}\right) 
	\left[-2 \eta \bar{w} f_{1}(\bar{w}) \right]\ln \Lambda
	\nonumber\\
	&+ 2 \pi g_{A} v_{F}^2 \left(\sih{\alphab}\kah{\betab}\otimes\sih{\alphab}\kah{\betab}\right)
	\left[2 \eta \bar{w} f_{2}(\bar{w}) \right]\ln \Lambda
	\nonumber\\
	&+ 2 \pi g_{A3} v_{F}^2 \left(\sih{\alphab}\kah{3}\otimes\sih{\alphab}\kah{3}\right)
	\left[2 \eta \bar{w} f_{2}(\bar{w}) \right]\ln \Lambda
	\nonumber\\
	&+ 2 \pi g_{m} v_{F}^2 \left(\sih{3}\kah{\betab}\otimes\sih{3}\kah{\betab}\right) 
	\left[
	\begin{aligned}
	&2 \eta \bar{w} f_{2}(\bar{w}) 
	\\
	&+ 2 \eta \bar{w} f_{3}(\bar{w})
	\end{aligned}
	\right]\ln \Lambda
	\nonumber\\
	&+ 2 \pi g_{v} v_{F}^2 \left(\sih{3}\kah{3}\otimes\sih{3}\kah{3}\right)
	\left[
	\begin{aligned}
	&2 \eta \bar{w} f_{2}(\bar{w}) 
	\\
	&+ 2 \eta \bar{w} f_{3}(\bar{w}) 
	\end{aligned}
	\right]\ln \Lambda.
\end{align}
The functions $f_{1}(\bar{w})$--$f_{3}(\bar{w})$ were defined by Eqs.~(\ref{f1f2Def}) and
(\ref{f3Def}).

The remaining diagrams pictured in Figs.~\ref{FigD14} and \ref{FigD15} involve 3-electron loops.
We consider first the ring diagrams in Fig.~\ref{FigD14}, beginning with $\mathsf{D}_{12.1}$:
\begin{align}\label{D14.1}
	\mathsf{D}_{12.1}
	=&
	2\pi \, \mathcal{G}^{\nu}_{\mu} v_{F}^2 2\pi \, \mathcal{G}^{\gamma}_{\lambda} v_{F}^2 
	(\sih{\lambda}\kah{\gamma})_{c,d} \left(i\sqrt{\frac{w}{N}}\right)^{2}\frac{-1}{1+\bar{w}}
	\nonumber\\
	&
	\begin{aligned}[b]
	\times\int
	&\frac{d\omega \, \dvex{l} \, \dvex{l'}}{(2\pi)^{5}} \,
	\frac{\left[\sih{\mu}\kah{\nu} v_{F} \sigmah\cdot\vex{l'}\right]_{a,b}}{v_{F}^2 \vex{l'}^{2}}
	\frac{1}{|\vex{l'}|}
	\\
	\times&
	\frac{\Tr
	\left\lgroup
	\begin{aligned}
		&\sih{\mu}\kah{\nu}[i \omega + v_{F} \sigmah\cdot(\vex{l}-\vex{l'})]
		\\
		&\times
		[i \omega + v_{F} \sigmah\cdot\vex{l}]
		\sih{\lambda}\kah{\gamma}
		\\
		&\times
		[i \omega + v_{F} \sigmah\cdot\vex{l}]
	\end{aligned}
	\right\rgroup}
	{[\omega^{2} + v_{F}^2 \vex{l}^{2}]^{2}[\omega^{2} + v_{F}^2 (\vex{l} - \vex{l'})^{2}]}.
	\end{aligned}
\end{align}
Simplifying, we obtain
\begin{align}\label{D14.1--II}
\mathsf{D}_{12.1}
	=&(2\pi)^{2} \mathcal{G}^{\nu}_{\mu}\mathcal{G}^{\nu}_{\lambda}
	v_{F}^2
	(\sih{\mu}\kah{\nu}\sih{\alphab})_{a,b}(\sih{\lambda}\kah{\nu})_{c,d} 
	\frac{2^{5} i \bar{w}}{1+\bar{w}}
	\nonumber\\
	&\times
	\left[ 
	\begin{aligned}
	&\epsilon_{\mu \lambda \betab} \, K_{1}^{\alphab,\betab}
	+
	\delta_{\betab,\xib} \epsilon_{\mu \lambda \etab}
	K_{2}^{\alphab,\betab,\etab,\xib}
	\\
	&-
	\left( 
	\delta_{\mu,\xib} \epsilon_{\lambda \etab \betab}
	+\delta_{\lambda,\xib} \epsilon_{\mu \etab \betab}
	\right)
	K_{2}^{\alphab,\betab,\etab,\xib}
	\end{aligned}
	\right],
\end{align}
where
\begin{align}\label{K1Def}
	K_{1}^{\alphab,\betab} &\equiv 
	\int 
	\frac{d\omega \, \dvex{l} \, \dvex{l'}}{(2\pi)^{5}} \,
	\frac{l'^{\alphab}(\vex{l} - \vex{l'})^{\betab}\omega^{2}}
	{|\vex{l'}|^{3}[\omega^{2} + \vex{l}^{2}]^{2}[\omega^{2} + (\vex{l} - \vex{l'})^{2}]},
	\nonumber\\
	&\sim
	- \frac{12}{2^{10}} \frac{\ln\Lambda}{2\pi} \delta^{\alphab,\betab},
\end{align}
and
\begin{align}\label{K2Def}
	K_{2}^{\alphab,\betab,\etab,\xib} &\equiv 
	\int 
	\frac{d\omega \, \dvex{l} \, \dvex{l'}}{(2\pi)^{5}} \,
	\frac{l'^{\alphab}(\vex{l} - \vex{l'})^{\betab} l^{\etab} l^{\xib}}
	{|\vex{l'}|^{3}[\omega^{2} + \vex{l}^{2}]^{2}[\omega^{2} + (\vex{l} - \vex{l'})^{2}]}.
	\nonumber\\
	&\sim
	\frac{1}{2^{10}}
	\frac{\ln\Lambda}{2\pi}
	\left[
	\begin{aligned}
	&3(\delta^{\alphab,\etab} \delta^{\betab,\xib} + \delta^{\alphab,\xib} \delta^{\betab,\etab})
	\\
	&-
	13(\delta^{\alphab,\betab} \delta^{\etab,\xib})
	\end{aligned}
	\right].
\end{align}
Eqs.~(\ref{K1Def}) and (\ref{K2Def}) may be obtained by a) performing the
ultraviolet-convergent 3-electron loop momentum integral over $\vex{l}$ via 
Feynman parameters, b) evaluating the frequency integration over
the real line, and c) completing the final ultraviolet-divergent momentum
integral over $\vex{l}'$.

Combining Eqs.~(\ref{D14.1--II})--(\ref{K2Def}),
and carefully summing barred (unbarred) indices over $\{1,2\}$ 
($\{1,2,3\}$), we finally obtain
\begin{align}\label{D14.1--III}
	\mathsf{D}_{12.1}
	&\sim
	\frac{v_{F}^2 \bar{w}}{1+\bar{w}}\frac{\ln\Lambda}{2}
	\left\lgroup
	\begin{aligned}
	&2 \pi \left(\sih{3}\kah{\betab}\otimes\sih{3}\kah{\betab}\right) 
	\left[-4 g_{m} g_{A} \right]
	\\
	&+ 2 \pi \left(\sih{3}\kah{3}\otimes\sih{3}\kah{3}\right)
	\left[-4 g_{v} g_{A3} \right]
	\end{aligned}
	\right\rgroup.
\end{align}
The remaining diagrams in Fig.~\ref{FigD14} give identical
contributions.
Next, we consider the diagrams shown in Fig.~\ref{FigD15}. Each of the diagrams 
$\mathsf{D}_{13.1}$--$\mathsf{D}_{13.4}$ represents an autorenormalization of the bare 
scalar potential disorder parameter $g_{u}$, mediated by the Coulomb interaction $\bar{w}$. 
Due to the 3-ring cancelation property encapsulated by Fig.~\ref{FigD9}, the sum 
of these diagrams gives (exactly) zero.

The bare vertex function is
\begin{align}\label{DisorderVertexBareCorrected}
	i \Gamma_{D}^{(0)} 
	=& 
	2 \pi g_{u} v_{F}^2 \left(\hat{1}\otimes\hat{1}\right)
	\nonumber\\
	&+ 2 \pi \mathcal{G}^{\nu}_{\mu} v_{F}^2 (\sih{\mu}\kah{\nu})\otimes(\sih{\mu}\kah{\nu})
	\nonumber\\
	&+ \mathsf{D}_{11} + \mathsf{D}_{12},
\end{align}	
where $\mathsf{D}_{m}$, $m \in \{11,12\}$, denotes the sum of all diagrams in Fig.~$m$.

The renormalization condition is
\begin{equation}
	\left[\dlambda + 2\frac{d \ln Z_{\psi}}{d\ln\Lambda}\right]i \Gamma_{D}^{(0)} = 0,
\end{equation}
which leads to the renormalization group flow equations
\begin{subequations}\label{GRG}
\begin{align}
	\frac{d g_{u}}{dl}
	=&
	2 g_{u}
	\left[
	\bGs - \eta \bar{w} f_{3}(\bar{w}) 
	\right]
	+ 8 g_{m} g_{A} + 4 g_{v} g_{A3},
	\label{guBareFlow}
	\\
	\frac{d g_{A}}{dl}
	=&
	8 g_{A} g_{A3} + 2 g_{m} (\tgu+g_{v}),
	\\
	\frac{d g_{A3}}{dl} 
	=&
	2 g_{m}^{2} + 8 g_{A}^{2} + 2\tgu g_{v},
	\\
	\frac{d g_{m}}{dl} 
	=&
	2 g_{m}(g_{v}-\tgu) + 4 g_{A} (\tgu + g_{v})
	+2 g_{m} \eta \bar{w} f_{3}(\bar{w})
	\nonumber\\
	&-\frac{16 \bar{w}}{1+\bar{w}} g_{m} g_{A},
	\\
	\frac{d g_{v}}{dl} 
	=&
	2 g_{v}(2 g_{m} - \tgu - g_{v}) + 4 g_{A3} (\tgu + g_{v}) 
	\nonumber\\
	&+ 8 g_{A} (g_{m} - g_{v}) 
	+2 g_{v} \eta \bar{w} f_{3}(\bar{w})
	-\frac{16 \bar{w}}{1+\bar{w}} g_{v} g_{A3},
\end{align}	
\end{subequations}
where we have used Eqs.~(\ref{f3Def}), Eqs.~(\ref{ZPsiRG}), and (\ref{wbarRG}).
[$\bGs$ was defined by Eq.~(\ref{bGsDef}).]


\section{Analysis of the renormalization group flows\label{Results}}

In this section, we use the results of our one-loop renormalization group (RG) study, detailed in 
Sec.~\ref{PerturbativeCalc}, to attempt to understand the topology of graphene's phase diagram in 
disorder--interaction coupling strength space. Our primary goal is to determine \emph{what kind} 
of low-energy theory one should use to understand the macroscopic electronic properties of a graphene 
sheet, using the RG as our guide. If the RG reveals a theoretically tractable (e.g., perturbatively 
accessible) phase or critical point, then we can use that framework to make predictions that can be tested
against experiment. As we will see, the situation appears to be more complex (at least to lowest order 
in $1/N$), and the ultimate low-energy physics (at zero temperature) likely requires a description 
very different from the weakly-perturbed, massless Dirac electron picture used as our starting point here. 

Nevertheless, the RG does allow us to identify several possible scaling regimes; our most
interesting result concerns the apparent robustness of one such scaling regime, at least as viewed
from the vantage point of the weakly-disordered Dirac electron theory. This (crossover) regime turns 
out to be dominated by the non-Abelian vector potential disorder, characterized by the parameters $g_{A}$ and $g_{A3}$ 
[Eqs.~(\ref{DisDef}) and (\ref{DisVar})] which appear, e.g., 
in the description of elastic lattice
deformations (``ripples''),\cite{MorozovNovoselov,MeyerGeim,MorpurgoGuinea,FasolinoLosKatsnelson} as well as
topological defects in the graphene lattice (Sec.~\ref{ClassCIandBDI} and 
Refs.~\onlinecite{TopologicalDisorder,OstrovskyGornyiMirlin1}). 
This principal result is demonstrated 
via numerical integration of the RG flows, and is discussed in subsection~\ref{GrapheneResults}, below. 
Potential experimental manifestations of our results are 
discussed in Sec.~\ref{Physics}.

\subsection{One-loop flow equations}

Using renormalized perturbation theory, we obtained in the previous section the one-loop flow equations
for the coupling constants of the large-$N$ graphene field theory defined by Eq.~(\ref{Sbar2}). 
Six coupling strengths 
define that model: the Fermi velocity $v_{F}$, the dimensionless Coulomb interaction strength $\bar{w}$ 
[Eqs.~(\ref{wbar}) and (\ref{wbarrs})], the screened scalar potential disorder strength $\tgu$ [Eq.~(\ref{tguDef})], 
as well as $g_{A}$, $g_{A3}$, $g_{m}$, and $g_{v}$ [Eqs.~(\ref{H})--(\ref{DisVar})].
The latter four disorder parameters were encoded in the disorder metric $\mathcal{G}_{\mu}^{\nu}$,
defined by Eq.~(\ref{DisMetric}).
From Eqs.~(\ref{lnvFRG}), (\ref{wbarRG}), and (\ref{GRG}) obtained in Sec.~\ref{PerturbativeCalc}, 
the flow equations to the lowest nontrivial order 
in the small parameters $1/N$ and $\{\tgu,\mathcal{G}_{\mu}^{\nu}\}$ are given by
\begin{subequations}\label{FlowEqs}
\begin{eqnarray}
	\frac{d \ln v_{F}}{dl} &=& z - 1 - \bGs + \eta \, \bar{w} f_{3}(\bar{w}), \label{lnvFFlow}
	\\
	\frac{d \ln \bar{w}}{dl} &=& \bGs - \eta \, \bar{w} f_{3}(\bar{w}), \label{wlnFlow}
	\\
	\frac{d \tgu}{dl}
	&=&
	\frac{2 \tgu}{1 + \bar{w}}
	\left[
	\bGs - \eta \bar{w} f_{3}(\bar{w}) 
	\right]
	+ \frac{8 g_{m} g_{A} + 4 g_{v} g_{A3}}{(1 + \bar{w})^{2}},
	\nonumber\\
	\label{guFlow}\\
	\frac{d g_{A}}{dl}
	&=&
	8 g_{A} g_{A3} + 2 g_{m} (\tgu+g_{v}),
	\label{gAFlow}\\
	\frac{d g_{A3}}{dl} 
	&=&
	2 g_{m}^{2} + 8 g_{A}^{2} + 2\tgu g_{v},
	\label{gA3Flow}\\
	\frac{d g_{m}}{dl} 
	&=&
	2 g_{m}(g_{v}-\tgu) + 4 g_{A} (\tgu + g_{v})
	+2 g_{m} \eta \bar{w} f_{3}(\bar{w})
	\nonumber\\
	&&-\frac{16 \bar{w} \, g_{m} g_{A}}{1+\bar{w}},
	\label{gmFlow}\\
	\frac{d g_{v}}{dl} 
	&=&
	2 g_{v}(2 g_{m} - \tgu - g_{v}) + 4 g_{A3} (\tgu + g_{v}) 
	\nonumber\\
	&&+ 8 g_{A} (g_{m} - g_{v}) 
	+2 g_{v} \eta \bar{w} f_{3}(\bar{w})
	-\frac{16 \bar{w} \, g_{v} g_{A3}}{1+\bar{w}}.
	\nonumber\\
	\label{gvFlow}
\end{eqnarray}	
\end{subequations}
In these equations,
\begin{equation}\label{bGsDef2}
	\bGs \equiv \tgu + 4 g_{A} + 2 g_{A3} + 2 g_{m} + g_{v},
\end{equation}
and
\begin{equation}\label{etaDef2}
	\eta \equiv \frac{8}{\pi N}.
\end{equation}

\begin{figure}
\includegraphics[width=0.4\textwidth]{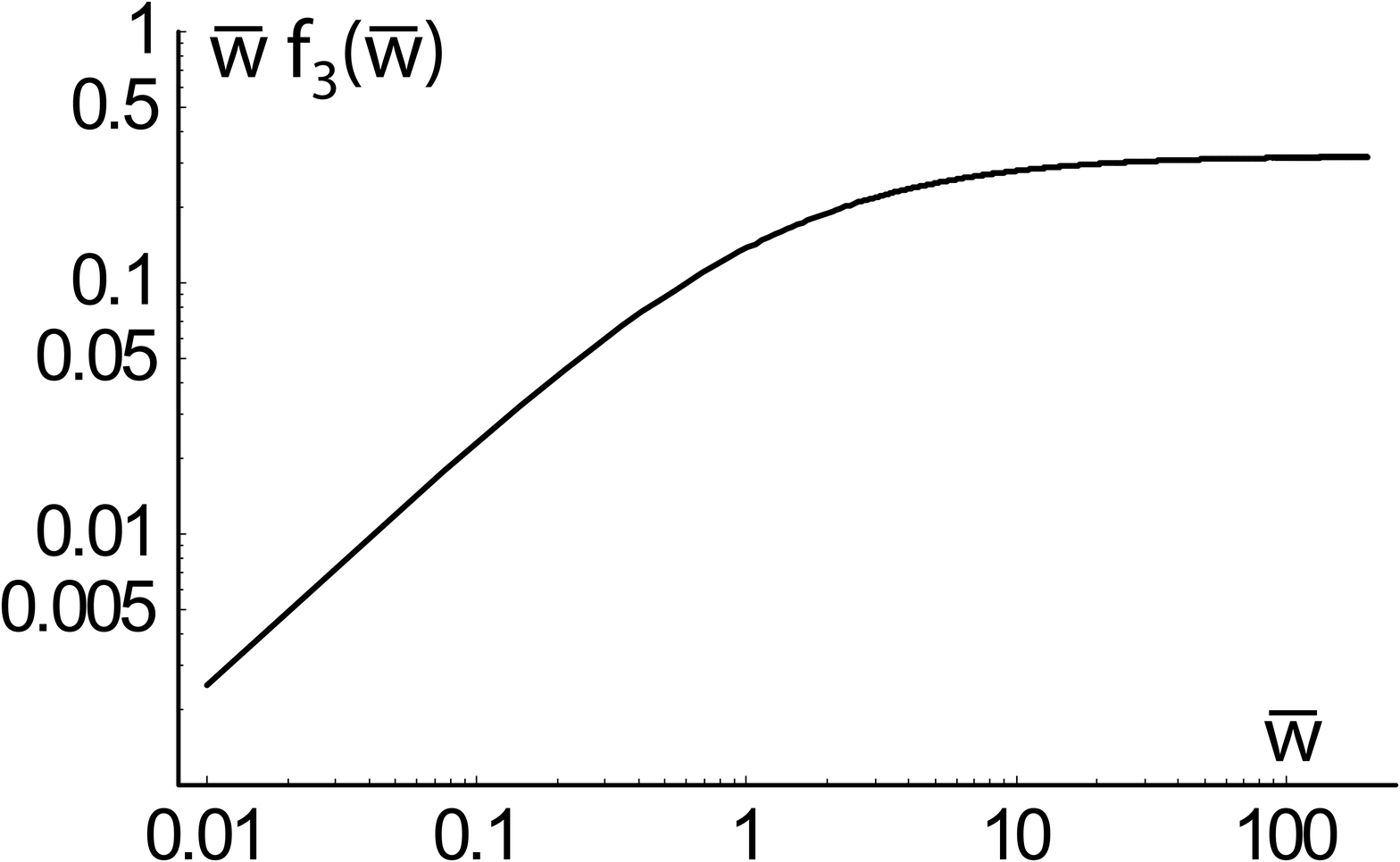}
\caption{Plot of $\bar{w} f_{3}(\bar{w})$.\label{fPlot}}
\end{figure}

In Eq.~(\ref{lnvFFlow}), $z = z(l)$ is the (as yet unspecified) dynamic critical exponent.
The function $f_{3}(\bar{w})$ appearing in Eq.~(\ref{FlowEqs}) was defined by Eq.~(\ref{f3Def}), 
repeated below for convenience:
\begin{align}\label{f3DefRepeat<}
	f_{3}(\bar{w})
	=&
	\frac{1}{2\bw^{2}}
	\left\lgroup
	\frac{2}{\pi}
	\left[
	\bar{w} + \frac{\arccos(\bar{w})}{\sqrt{1-\bar{w}^{2}}}
	\right]
	 - 1 
	\right\rgroup,
\end{align}
valid for $\bar{w} \leq 1$. The analytical continuation of $f_{3}(\bar{w})$ 
for $\bar{w} \geq 1$ is given by
\begin{align}\label{f3DefRepeat>}
	f_{3}(\bar{w})
	=&
	\frac{1}{2 \bar{w}^{2}}
	\left\lgroup
	\begin{aligned}
	&
	- 1
	+\frac{2 \bar{w}}{\pi}
	\\
	&
	+\frac{2}{\pi\sqrt{\bar{w}^{2}-1}} \ln\left(\bar{w} + \sqrt{\bar{w}^{2}-1}\right)
	\end{aligned}
	\right\rgroup.
\end{align}
In the small and large $\bar{w}$ limits, $f_{3}(\bar{w})$ has the behaviors
\begin{align}\label{f3Lim}
	f_{3}(\bar{w})
	&\sim \frac{1}{4} - \frac{2 \bar{w}}{3 \pi} + \frac{3 \bar{w}^{2}}{16} + \ldots
	&& 0 \leq \bar{w} \ll 1,
	\nonumber\\
	&\sim \frac{1}{\pi \bar{w}} + \ord{\frac{1}{\bar{w}^{2}}},
	&& \bar{w} \gg 1.
\end{align}
The function $\bar{w} f_{3}(\bar{w})$ is plotted in Fig.~\ref{fPlot}.

Eq.~(\ref{guFlow}) follows from the definition of the screened scalar potential disorder
strength $\tgu$ [Eq.~(\ref{tguDef})], and the flow equation (\ref{guBareFlow}) for
the bare scalar potential strength. We emphasize that it is $\tgu$ that appears in
physical quantities in the large-$N$ theory, as demonstrated in 
Sec.~\ref{Physics};
the renormalization group must therefore be formulated in terms of $\tgu$, the 
effective scalar potential strength.

\subsection{Termination of the RG flow for finite temperatures and chemical potential}\label{sec:termination}

The derived RG equations \rref{FlowEqs} are limited by the regime of the weak
disorder strength and usually become non-usable when $l\to \infty$. 
There are, however, physical situations when those equations are
adequate to determine the transport properties of the system.
Those are regimes of either sufficiently large doping
(characterized by the chemical potential $\mu$), or of sufficiently
high temperature $T$. 
At such an energy scale, the logarithmic renormalizations must be stopped,
and kinetic coefficients computed, as
in Sec.~\ref{Physics}.

To use \reqs{FlowEqs} we have to relate the 
scale $l_*$ at which the RG is stopped to the 
energy scale 
\be
	\varepsilon_* \simeq {\rm max} \left(|\mu|,T\right). 
	\label{epsilon*}
\ee
To achieve this goal, we write a scaling equation
\begin{equation}\label{APP-TempScaling}
        \frac{d \ln \epsilon_*}{d l_*} = -z(l_*),
\end{equation}
and find the scale-dependent dynamic critical exponent 
by setting Eq.~(\ref{lnvFFlow}) equal to zero.\cite{BK,BDIFosterLudwig}
It gives us an implicit function $l_*(\epsilon_*)$
\be
	\label{l*}
	\begin{split}
	&\ln \frac{\epsilon_* a}{v_F^{(a)}}=- \int_0^{l_*}{dl}z(l);
	\\
	&z(l)= 1 + \bGs(l) - \eta\, \bar{w}(l) f_{3}[\bar{w}(l)],
	\end{split}
\ee
where $v_F^{(a)}$ is the Fermi velocity defined on the spatial
scale of the order of lattice constant $a$.

\subsection{Restricted flow equations: SU(2) valley space rotation symmetry restored}

A useful subset of the RG flow provided by Eq.~(\ref{FlowEqs})
is to consider the system with SU(2) valley ($\kappa$) 
space rotational symmetry restored \emph{on average}. Such a restriction constrains
\begin{align}\label{SU(2)VSDDef}
	g_{A} &= g_{A3}, \nonumber\\
	g_{m} &= g_{v}.
\end{align}
[See Eqs.~(\ref{DisDef}) and (\ref{DisVar}).]
Imposing Eq.~(\ref{SU(2)VSDDef}), the RG equations reduce to
\begin{subequations}\label{FlowEqsVSD}
\begin{eqnarray}	
	\frac{d \ln v_{F}}{dl} 
	&=& 
	z - 1  
	- \left(\tgu+ 6 g_{A} + 3 g_{m}\right)
	+ \eta \, \bar{w} f_{3}(\bar{w}), 
	\label{lnvFFlowVSD}\\
	\frac{d \ln \bar{w}}{dl} 
	&=& 
	\left(\tgu+ 6 g_{A} + 3 g_{m}\right) - \eta \, \bar{w} f_{3}(\bar{w}),
	\label{lnwFlowVSD}\\
 	\frac{d \tgu}{d l}
	&=&
	\frac{2 \tgu}{1 + \bar{w}}
	\left[
	\tgu + 6 g_{A} + 3 g_{m} - \eta \, \bar{w} f_{3}(\bar{w})
	\right]
	\nonumber\\
	&&+\frac{12 g_{m} g_{A}}{(1 + \bar{w})^{2}},			
	\label{guFlowVSD}
	\\
	\frac{d g_{A}}{d l}
	&=&
	8 g_{A}^{2} + 2 g_{m}(\tgu + g_{m}),
	\label{gAFlowVSD}
	\\
	\frac{d g_{m}}{d l} 
	&=&
	2 g_{m}(g_{m}-\tgu) + 4 g_{A} \tgu
	+4\left(\frac{1-3\bar{w}}{1+\bar{w}}\right) g_{m} g_{A}
	\nonumber\\
	&&+2 g_{m} \eta \bar{w} f_{3}(\bar{w}).
	\label{gmFlowVSD}
\end{eqnarray}	
\end{subequations}
As discussed below in Sec.~\ref{GrapheneResults}, statistical SU(2) Fermi
space ($\kappa$) rotational symmetry is \emph{generically} 
restored under integration of the full flow equations, Eq.~(\ref{FlowEqs}); 
we will therefore focus upon the restricted flow equations (\ref{FlowEqsVSD}) when we 
discuss our primary results in Sec.~\ref{GrapheneResults}.

Note that Eq.~(\ref{gAFlowVSD}) implies that the only non-trivial fixed point 
structure (a fixed line) occurs in these restricted equations when $g_{A} = g_{m} = 0$
(since the disorder parameters, being variances [Eq.~(\ref{DisVar})], cannot 
take negative values).

\subsection{Graphene with non-generic disorder}

Before analyzing graphene subject to generic (short-range correlated), time-reversal invariant disorder, 
we discuss several simpler limiting cases that arise upon fine-tuning the disorder distribution defined 
by Eqs.~(\ref{DisDef}) and (\ref{DisVar}). To do so, we specialize the flow Eq.~(\ref{FlowEqs}) to the 
quantum disorder universality classes discussed in subsections \ref{ClassAII}--\ref{ClassD}. Although it is 
typically difficult to fine-tune the disorder profile experimentally, some of the limiting cases discussed 
below may dominate various scaling or crossover regimes in graphene.
(See also Sec.~\ref{Physics}.)

\subsubsection{Scalar potential disorder and the sympletic class AII}\label{ClassAIIRG}

If remote charged impurities provide the principal scattering mechanism in the experiments discussed
in Refs.~\onlinecite{Geim1,Geim2,Kim,Kim2}, 
one would expect fluctuations in the potential $u(\vex{r})$ to dominate over the other disorder types.
As a zeroth order approximation, we may neglect the other disorder potentials altogether, setting
$g_{A} = g_{A3} = g_{m} = g_{v} = 0$ [Eqs.~(\ref{DisDef}) and (\ref{DisVar})]. 
As discussed in Sec.~\ref{ClassAII}, this is equivalent to enforcing $\textrm{SU}(2)$ 
valley space rotational symmetry [invariance of Eq.~(\ref{DisDef}) under $\kappa$-space rotations] 
\emph{in every realization of the static disorder}. This theory satisfies the effective $\textrm{TRI}^{*}$ 
condition, defined by Eq.~(\ref{TRIstar}), and belongs to the symplectic (``spin-orbit'') 
ordinary metal class AII.\cite{AleinerEfetov,OstrovskyGornyiMirlin1,OstrovskyGornyiMirlin2,
BardarsonTworzydloBrouwerBeenakker,NomuraKoshinoRyu,RyuMudryObuseFurusaki} 
The flow equations are
\begin{subequations}\label{FlowAII}
\begin{align}	
	\frac{d \ln \bar{w}}{dl} 
		=& 
		\tgu -\eta \, \bar{w} f_{3}(\bar{w}), 
	\label{lnwFlowAII}
	\\
	\frac{d \tgu}{d l}
		=&
		\frac{2 \tgu}{1 + \bar{w}}
		\Big[
		\tgu - \eta \bar{w} f_{3}(\bar{w})
		\Big].			
	\label{guFlowAII}
\end{align}	
\end{subequations}
These equations possess an unstable fixed line for $\bg{u} = \eta \bar{w} f_{3}(\bar{w})$; moreover,
\begin{equation}\label{ParaRG}
	g_{u}(l) = C \bar{w}^{2}(l),
\end{equation}
where $C = g_{u}(0)/\bar{w}^{2}(0)$, so that the RG flows are parabolic trajectories
in the $g_{u}-\bar{w}$ plane. [$g_{u} = \tgu(1+\bar{w})^{2}$ is the unscreened scalar
potential disorder strength; see Eq.~(\ref{tguDef}).]

A similar result was found previously for the case of graphene at weak Coulomb interaction 
coupling;\cite{Ye,StauberGuineaVozmediano} in the case of the large-$N$ generalization, 
a technical error in Ref.~\onlinecite{Ye} (see Sec.~\ref{SLFNRG} and Appendix~\ref{APP-PsiSLFNRG}) led to the 
prediction that the line parameterized by Eq.~(\ref{ParaRG}) becomes attractive at larger Coulomb 
interaction strengths. Instead, Eq.~(\ref{FlowAII}), coupled with the monotonic behavior of $\bar{w} f_{3}(\bar{w})$,
[Eqs.~(\ref{f3DefRepeat<})--(\ref{f3Lim}) and Fig.~\ref{fPlot}], shows that this line is repulsive for all 
$\bar{w} \geq 0$. In other words, for the case of scalar potential disorder only, the physics 
in both the weak coupling and large-$N$ pictures is qualitatively the same.
The stability of the $\tgu$-$\bar{w}$ fixed line in the full 6-dimensional disorder-interaction coupling
constant space is analyzed in Appendix~\ref{APP-FixedLines}.

\subsubsection{Particle-hole symmetry and the chiral class BDI}\label{ClassBDIRG}

As discussed below Eq.~(\ref{DisVar}) in Sec.~\ref{LatticeModelandSym}, different quantum disorder (random matrix)
universality classes may be theoretically realized by enforcing invariance of the graphene system, in every
realization of disorder, under different combinations of the transformations defined by Eqs.~(\ref{PH})--(\ref{TRIstar}).
If we enforce both particle-hole symmetry (PH) and time-reversal invariance (TRI) [Eqs.~(\ref{PH}) and (\ref{TRI})]
in every static disorder realization, consistent with, e.g., the presence of carbon atom vacancies 
(treated as scattering centers in the unitary, hard-scattering limit) and
the absence of further-neighbor hopping,\cite{footnote-d}
then the system falls into the ``chiral orthogonal'' class 
BDI.\cite{BernardLeClair,AltlandSimonsZirnbauer,OstrovskyGornyiMirlin1,HatsugaiWenKohmoto,GuruswamyLeClairLudwig,RyuHatsugai} 
The allowed disorder strengths are $g_{m}$ and $g_{A3}$, 
and the one-loop flow equations are 
\begin{subequations}\label{FlowBDI}
\begin{align}	
	\frac{d \ln \bar{w}}{dl} 
		=& 
		2 g_{A3} + 2 g_{m} - \eta \, \bar{w} f_{3}(\bar{w}), 
	\label{lnwFlowBDI}
	\\
	\frac{d g_{A3}}{d l} 
		=& 
		2 g_{m}^{2},
	\label{gAFlowBDI}
	\\
	\frac{d g_{m}}{d l} 
		=& 
		2 g_{m} \eta \bar{w} f_{3}(\bar{w}).
	\label{gmFlowBDI}
\end{align}	
\end{subequations}
These equations possess no fixed point for $g_{m} \neq 0$, and generically flow to both strong 
disorder ($g_{A3},g_{m} \rightarrow \infty$) and strong interaction ($\bar{w}\rightarrow\infty$)
coupling. In passing, we note that, in the absence of Coulomb interactions ($\bar{w} = 0$), 
Eqs.~(\ref{FlowBDI}) can be extended to all orders in $g_{m}$ and $g_{A3}$, using conformal field 
theory methods.\cite{GuruswamyLeClairLudwig} One then finds that disordered graphene with strict 
particle-hole symmetry [Eq.~(\ref{PH})] and no long- or short-ranged interparticle interactions\cite{BDIFosterLudwig} 
possesses a critical, delocalized phase, indicative of the existence of extended single-particle 
states at zero energy, for arbitrary strength disorder. 

The theory with $g_{m} = 0$ does possess an attractive fixed line in the $g_{A3}$--$\bar{w}$ plane.
This line has been discussed before;\cite{Ye,StauberGuineaVozmediano} 
it is unlikely to play an important 
role in real graphene physics, because it is highly unstable in other disorder directions, as shown in 
Appendix~\ref{APP-FixedLines}.

\subsubsection{Vector potential disorder, ripples and topological defects: class CI}\label{ClassCIRG}

If we enforce both $\textrm{PH}^{*}$ and TRI [Eqs.~(\ref{PHstar}) and (\ref{TRI})], then the 
system falls into the CI quantum disorder class.\cite{BernardLeClair,AltlandSimonsZirnbauer,OstrovskyGornyiMirlin1} 
Recall from Sec.~\ref{LatticeModelandSym} that $\textrm{PH}^{*}$ describes an \emph{effective} 
particle-hole transformation, different from the physical PH transformation defined by Eq.~(\ref{PH}), 
the latter of which is inherited directly from the lattice model [Eq.~(\ref{H0})]. The allowed disorder 
strengths are $g_{A3}$ and $g_{A}$, i.e.\ the pure Dirac theory perturbed only by a quenched, 
$\textrm{SU}(2)$ non-Abelian vector potential. 

As discussed in Sec.~\ref{ClassCIandBDI}, the intravalley (Abelian) vector potential $\{A^{3}_{\alphab}\}$
appears in the description of long wavelength ``ripples'' in the low-energy Dirac theory.\cite{MorozovNovoselov,MeyerGeim,MorpurgoGuinea}
Both the intravalley and intervalley $\{A^{\betab}_{\alphab}\}$ vector potential components play a 
primary role in the field theoretic description of honeycomb lattice dislocations and disclinations.\cite{TopologicalDisorder}

The RG flow equations are
\begin{subequations}\label{FlowCI}
\begin{align}	
	\frac{d \ln \bar{w}}{dl} 
		=& 
		4 g_{A} + 2 g_{A3} - \eta \, \bar{w} f_{3}(\bar{w}), 
	\label{lnwFlowCI}
	\\
	\frac{d g_{A}}{d l} 
		=&
		8 g_{A} g_{A3},
	\label{gA2FlowCI}
	\\
	\frac{d g_{A3}}{d l} 
		=&
		8 g_{A}^{2}.
	\label{gAFlowCI}
\end{align}	
\end{subequations}
These equations possess no fixed point for $g_{A} \neq 0$, and generically flow to both strong disorder
and interaction coupling. In the absence of interactions, $\bar{w} = 0$, the Dirac theory with only non-Abelian
vector potential disorder flows to strong coupling; the flow terminates at an exactly-solvable conformal field theory fixed 
point\cite{NersesyanTsvelikWenger,CILogCFT}
[equivalent to the $\textrm{Sp}(2n)$ principal chiral non-linear sigma model, augmented with a WZW 
term].\cite{AltlandSimonsZirnbauer} Like the non-interacting BDI class discussed above, 
the fixed point of the non-interacting CI class model (with both $g_{A}$ and $g_{A3}$ nonzero) describes a critical, 
delocalized phase possessing extended wavefunctions near zero energy.\cite{NersesyanTsvelikWenger,CILogCFT}
This disorder class (in the absence of interactions) was discussed specifically in the context of graphene
in Ref.~\onlinecite{OstrovskyGornyiMirlin1}; these authors argued that the conductance of the non-interacting
CI class model should be of order the conductance quantum, and independent of the disorder strength.

\subsection{Graphene: generic disorder\label{GrapheneResults}}

At last, we turn to an analysis of our large-$N$ graphene field theory, characterized
by the full one-loop RG flow equations (\ref{FlowEqs}). In the absence of any fine-tuning
of the disorder potentials, our graphene model possesses time-reversal invariance
[Eq.~(\ref{TRI})] and (physical) spin $\textrm{SU}(2)$ rotational symmetry in every disorder 
realization. This system is in the standard orthogonal metal class AI. In the absence of 
interactions, all single particle wavefunctions are expected to be exponentially localized 
in 2D.\cite{LeeRamakrishnan}

\begin{figure}
\includegraphics[width=0.4\textwidth]{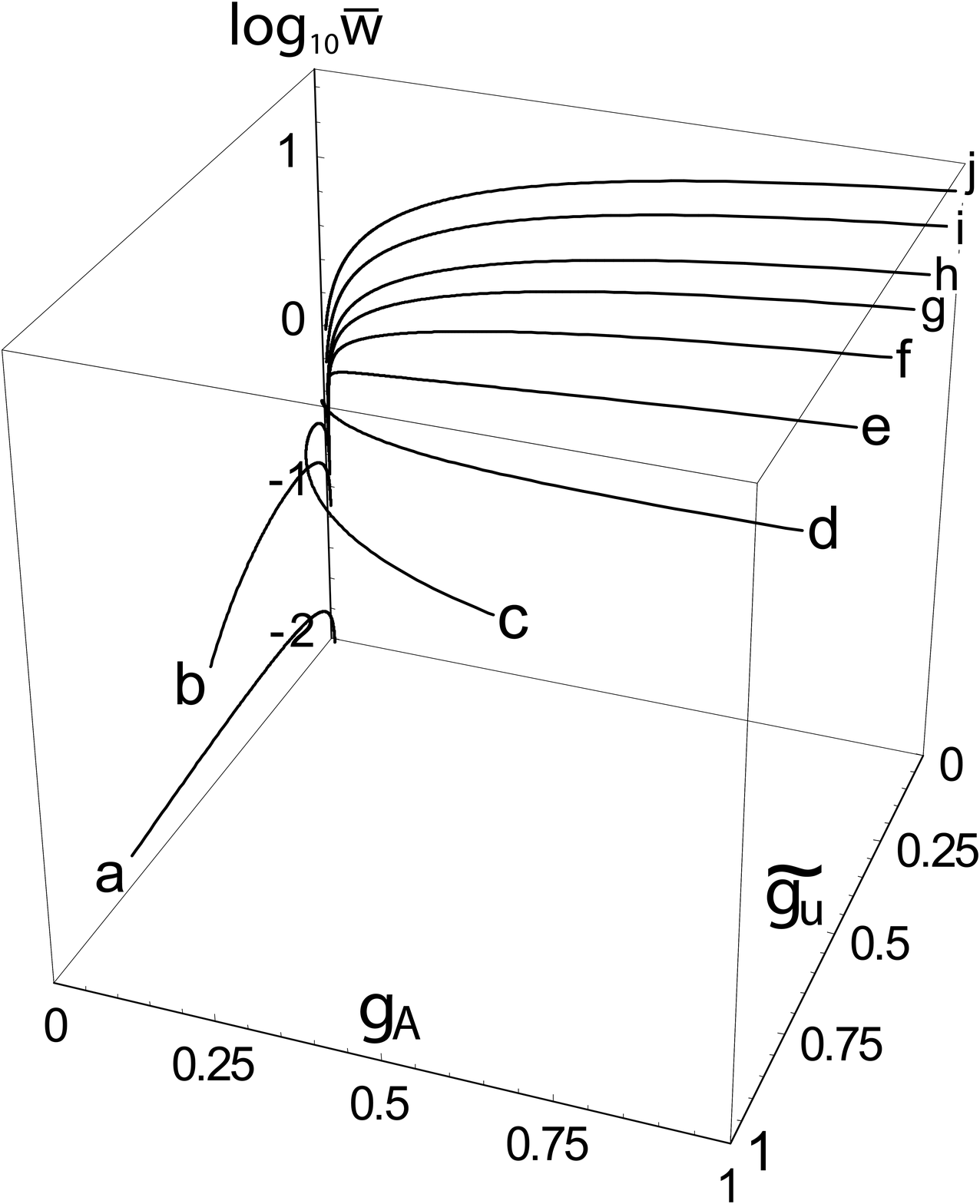}
\caption{
Set of 10 RG flow trajectories obtained via numerical integration for 
different initial Coulomb interaction strengths $\bar{w}$.
Here, we study the case with SU(2) valley space symmetry restored \emph{on average}, 
using Eq.~(\ref{FlowEqsVSD}). The initial conditions for the disorder 
coupling strengths are $\tgu(0) = g_{A}(0) = g_{m}(0) = 0.01$ for all trajectories
shown in the figure.
We have set the expansion parameter $\eta = 0.3$ [see Eq.~(\ref{etaDef2})];
the shape of the trajectories is only weakly dependent upon $\eta$ (within our
one-loop approximation).
Lower-case Latin letters label trajectories with different initial Coulomb interaction
strengths, given by
$\bar{w} =$
(a) 0.010, (b) 0.079, (c) 0.13, (d) 0.16, (e) 0.20, (f) 0.25, (g) 0.32, (h) 0.40,
(i) 0.63 (j) 1.0.
We do not exhibit the RG evolution of the random mass parameter $g_{m}$ 
[see Eqs.~(\ref{DisDef}), (\ref{DisVar}), (\ref{FlowEqsVSD})], because we find that 
its behavior is always subleading compared to the scalar ($\tgu$) or vector ($g_{A}$) 
potential parameters. 
\label{RGFlow1--Origin}}
\end{figure}

\begin{figure}
\includegraphics[width=0.4\textwidth]{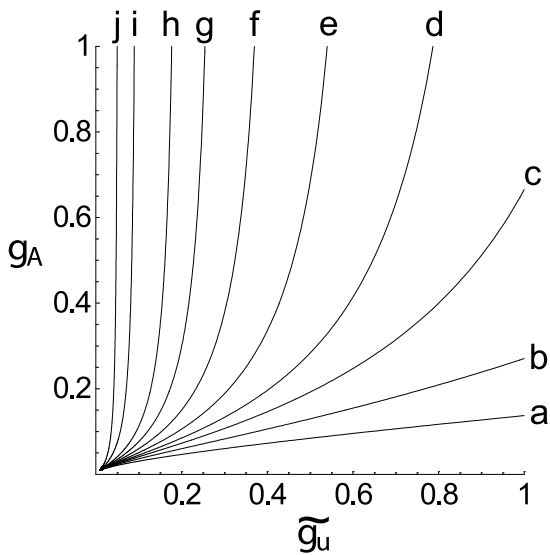}
\caption{Same RG trajectories shown in Fig.~\ref{RGFlow1--Origin}, but projected into the $\tgu-g_{A}$ disorder plane.
\label{RGFlow1--OriginElev}}
\end{figure}

The full flow Eqs.~(\ref{wlnFlow})--(\ref{gvFlow}) possess several fixed line structures
involving a \emph{single} nonzero disorder parameter and the Coulomb interaction strength $\bar{w}$.
The case of scalar potential disorder $\tgu$ was discussed above, in Sec.~\ref{ClassAIIRG}. 
Weak coupling analogs of this and other fixed lines were discussed in 
Refs.~\onlinecite{StauberGuineaVozmediano,Ye}, while large-$N$ versions were obtained and analyzed in 
Ref.~\onlinecite{Ye} [but see the discussion following Eq.~(\ref{ParaRG}), above].
We will not dwell upon these structures here, however, because
all of them are exceedingly unstable in the full 6-dimensional disorder-interaction coupling
constant space. (This fact is demonstrated explicitly in Appendix~\ref{APP-FixedLines}.)

The RG flow equations [Eq.~(\ref{FlowEqs})] possess no perturbatively accessible, 
isolated fixed points, other than the (trivial) non-interacting, clean (not disordered) Dirac fixed point,
defined by the condition 
\begin{equation}\label{FreeDiracFixedPoint}
	\bar{w} = \tgu = g_{A} = g_{A3} = g_{m} = g_{v} = 0.
\end{equation} 
In order to gleam information about graphene's low-energy physics (to order $1/N$ in our
large-$N$ calculational scheme), it is necessary to integrate the flow equations, 
and observe the evolution of the flow trajectories as a function of initial conditions 
in coupling constant space. In doing so, we seek to address two crucial questions
that so far remain unanswered:
\begin{enumerate}
	\item{Over what range of initial conditions, if any, does the theory flow back to the
		clean, non-interacting Dirac fixed point [Eq.~(\ref{FreeDiracFixedPoint})]. 
		In other words, can we find a critical manifold of codimension 0, with the Dirac 
		fixed point as its sink?}
	\item{If the clean, non-interacting Dirac fixed point is found to be unstable, where
		does the system flow to? What is the correct low-energy theory that we should use
		to understand disordered, interacting graphene?}
\end{enumerate} 
We can easily answer the first question, but we will be forced to speculate regarding the second.

We have numerically integrated Eqs.~(\ref{wlnFlow})--(\ref{gvFlow}) over a wide range of
disorder and interaction coupling strength initial conditions. In our one-loop approximation,
the resulting flow topology is only weakly dependent upon the expansion parameter $\eta$, defined by
Eq.~(\ref{etaDef2}). We find that the non-disordered, non-interacting 
Dirac fixed point, located by Eq.~(\ref{FreeDiracFixedPoint}), as well as the \emph{entirety} of the adjoining 
clean, but Coulomb-interacting line, $0 \leq \bar{w} < \infty$ and $\tgu = g_{A} = g_{A3} = g_{m} = g_{v} = 0$, 
is 
unstable in the presence 
of arbitrarily weak, but generic disorder [i.e.\ all five disorder strength parameters in 
Eq.~(\ref{FlowEqs}) nonzero, but arbitrarily small]. We stress that the entirety of the clean, interacting
line $\bar{w} \geq 0$ is perturbatively accessible in the $N \rightarrow \infty$ limit, 
and flows back to the non-interacting Dirac fixed point in the \emph{absence} of disorder.\cite{YeSachdev,Son}

\begin{figure}[b]
\includegraphics[width=0.4\textwidth]{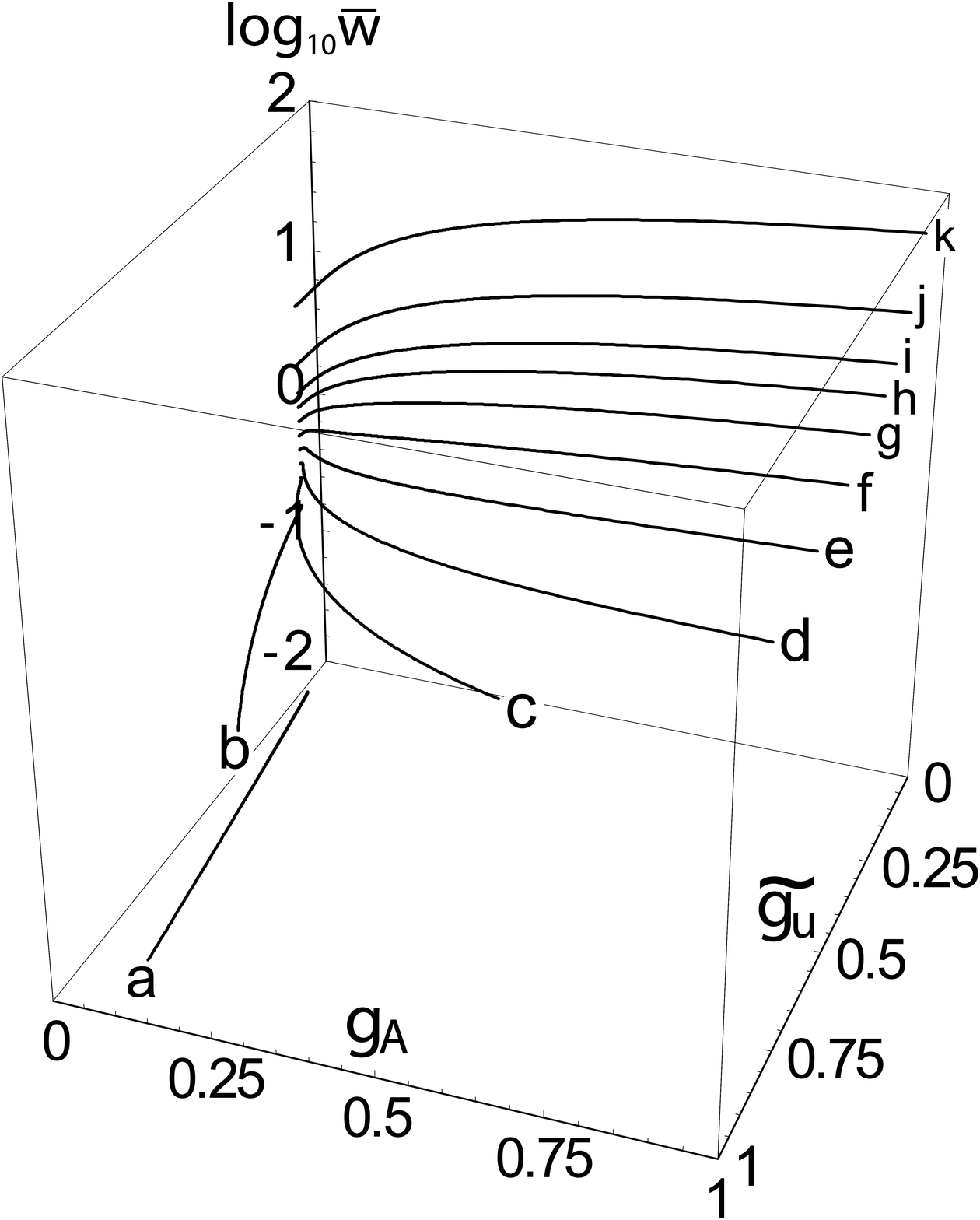}
\caption{As in Fig.~\ref{RGFlow1--Origin} but with the 
initial conditions for the disorder coupling strengths
$\tgu(0) = g_{m}(0) = 0.1$, $g_{A} = 0.01$.
Lower-case Latin letters label trajectories with different initial Coulomb interaction
strengths, given by
$\bar{w} =$
(a) 0.010, (b) 0.25, (c) 0.40, (d) 0.50, (e) 0.63, (f) 0.79, (g) 1.0, (h) 1.3, \
(i) 1.6, (j) 2.5, (k) 6.3.
\label{RGFlow2--gu}}
\end{figure}

\begin{figure}
\includegraphics[width=0.4\textwidth]{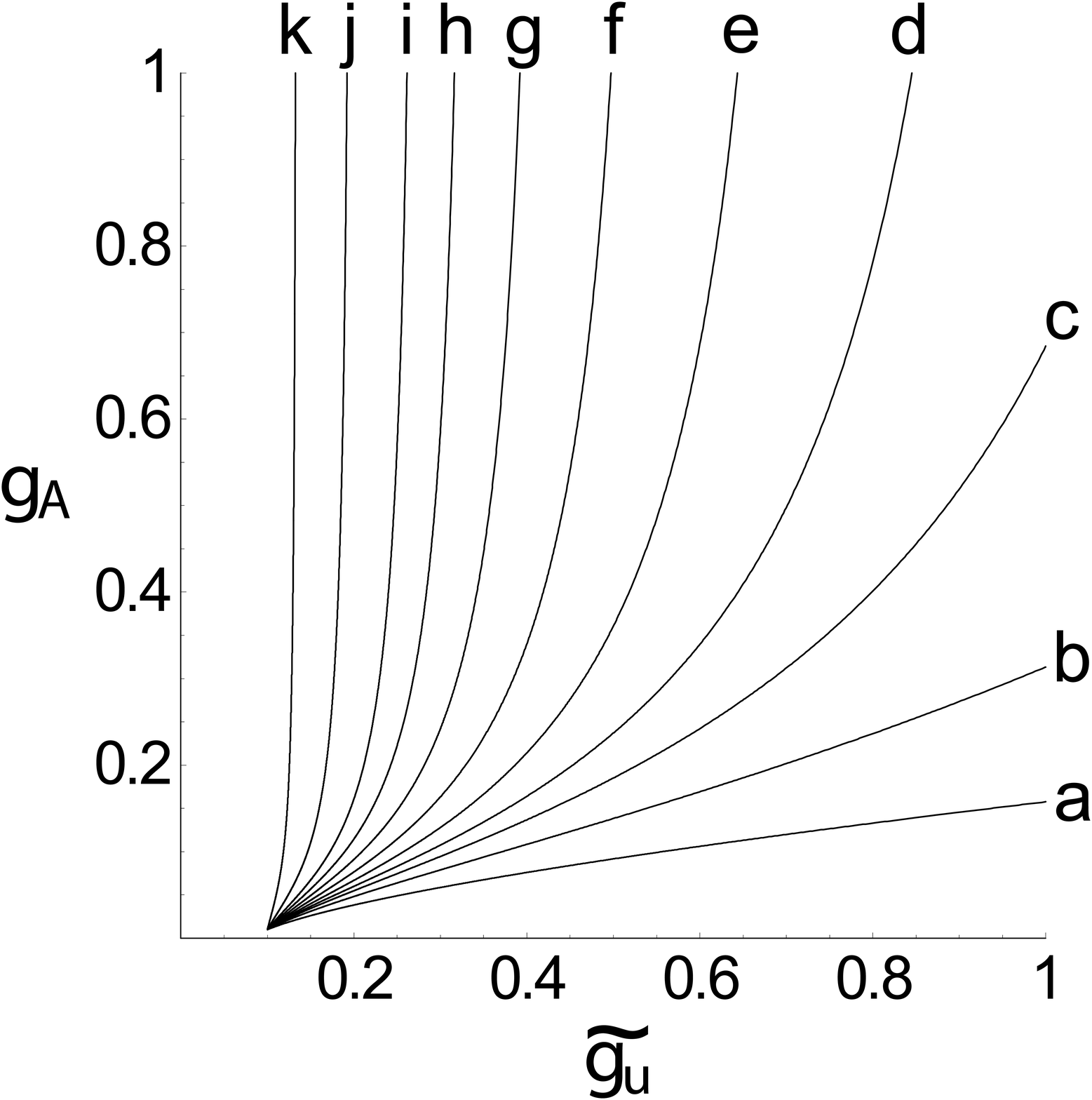}
\caption{Same RG trajectories shown in Fig.~\ref{RGFlow2--gu}, but projected into the $\tgu-g_{A}$ disorder plane.
\label{RGFlow2--guElev}}
\end{figure}

\begin{figure}[t]
\includegraphics[width=0.4\textwidth]{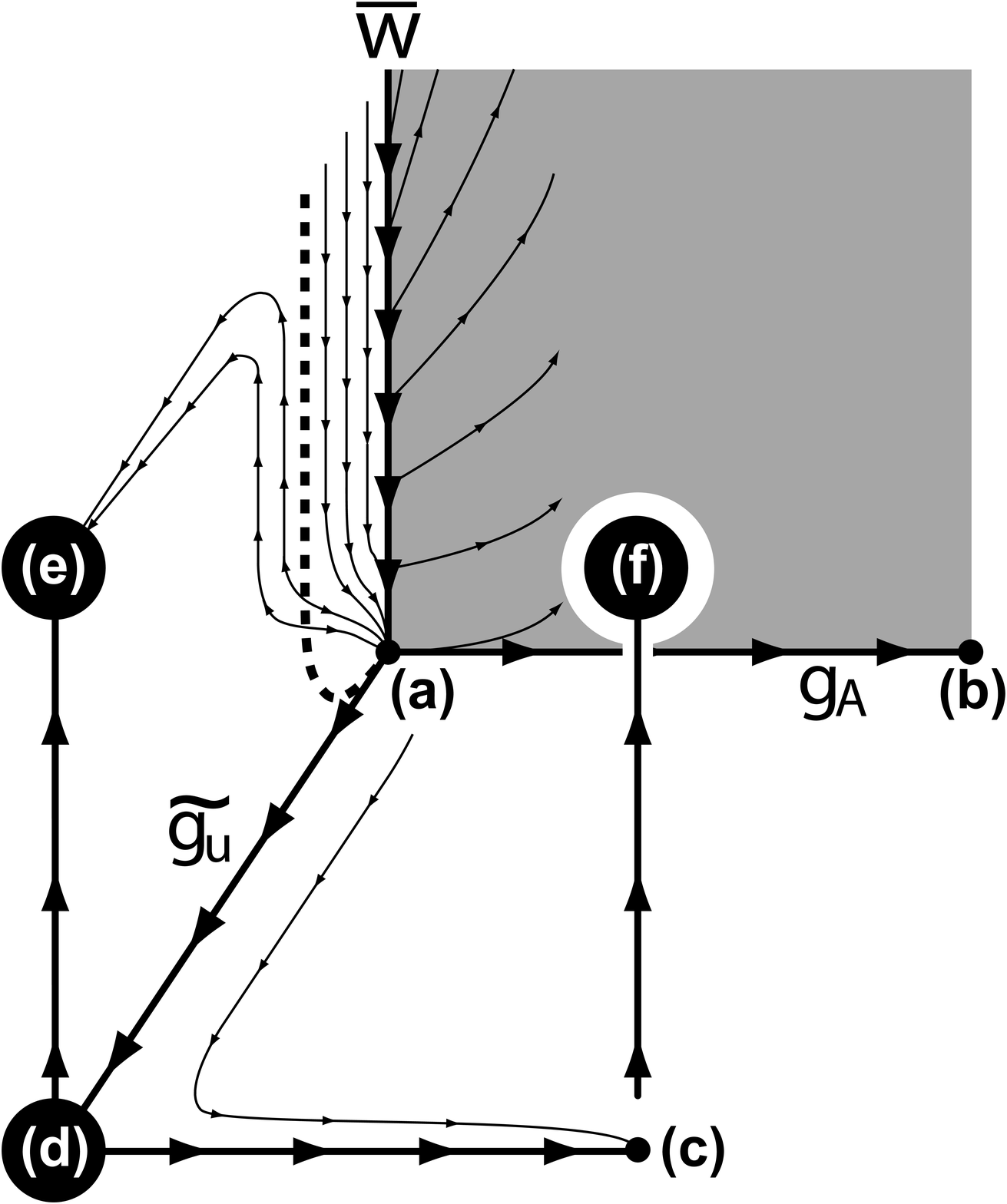}
\caption{
Schematic 
phase diagram for disordered, interacting graphene,
projected into the disorder ($\tgu$, $g_{A}$) interaction ($\bar{w}$)
coupling constant space, based in part upon the perturbative RG 
calculations performed at weak coupling,\cite{AleinerEfetov,StauberGuineaVozmediano}
and within the large-$N$ approximation (as detailed in this paper). The labeled 
small dots represent ``known'' phases, while the large black globes represent ``non-linear 
sigma model (NL$\sigma$M) arenas,'' wherein the system is described by a well-known (if poorly understood) 
model, which may possess multiple phases in $\textrm{d} = 2$ spatial dimensions.\cite{LeeRamakrishnan,BK} 
Point {\bf (a)} represents non-interacting, clean (not disordered) graphene, as located by 
Eq.~(\ref{FreeDiracFixedPoint}).
This theory is a sink for the clean, Coulomb interacting line (the vertical 
axis), as well as a portion of the 
$\tgu-\bar{w}$ plane. It is unstable to generic disorder perturbations. 
The dashed line in the $\tgu-\bar{w}$ is the unstable fixed line located
by Eqs.~(\ref{lnwFlowAII}) and (\ref{guFlowAII}).
The point {\bf (b)} 
is the non-interacting, non-Abelian SU(2) vector potential disorder 
CFT.\cite{NersesyanTsvelikWenger,CILogCFT}
Both points {\bf (a)} and {\bf (b)} are embedded in the particle-hole
symmetric plane $\tgu = g_{m} = 0$, represented by the shaded plane in the figure.
Point {\bf (c)} represents the non-interacting Anderson insulator for the orthogonal 
metal class (AI) in 2D. Globe {\bf (d)} should be described by the symplectic/spin-orbit 
class (AII), non-interacting NL$\sigma$M, 
possibly augmented by a topological 
term.\cite{AleinerEfetov,OstrovskyGornyiMirlin1,OstrovskyGornyiMirlin2,
BardarsonTworzydloBrouwerBeenakker,NomuraKoshinoRyu,RyuMudryObuseFurusaki}
In principle, both diffusive metallic and Anderson insulating phases are possible.\cite{LeeRamakrishnan} 
Globe {\bf (e)} represents 
the symplectic class Finkel'stein non-linear sigma model (FNL$\sigma$M) for interacting
electrons;\cite{CastellaniDiCastroForgacsSorella,KotliarSorella,BK} 
globe {\bf (f)} represents the orthogonal class (AI) 
FNL$\sigma$M.\cite{Finkelstein,CastellaniDiCastroLeeMaSorellaTabet,BK}
The numerical RG flows exhibited in Figs.~\ref{RGFlow1--Origin} and \ref{RGFlow2--gu}
should be ``superimposed'' upon this figure.
\label{PhaseDiag}}
\end{figure}

In the advent of nonzero quenched randomness, the flow to strong coupling exhibits two key features:
a) 
SU(2) valley space rotational symmetry is generically restored \emph{on average} upon integration of Eq.~(\ref{FlowEqs}) 
for general initial conditions, i.e.\ the differences $(|g_{A} - g_{A3}|)/(g_{A} + g_{A3})$ and 
$(|g_{m} - g_{v}|)/(g_{m} + g_{v})$ asymptote to zero as the solutions to the RG equations reach the 
limit of their validity.
From here on in, we will therefore restrict our discussion to the flows of the valley 
space SU(2)-symmetric system, described by Eq.~(\ref{FlowEqsVSD}).
b)		
The \emph{direction} of the diverging RG flows depends strongly upon the initial strength of the 
Coulomb interaction parameter $\bar{w}$. For weak (or vanishing) Coulomb interactions $\bar{w} \ll 1$,
the flow is dominated by the divergence of the screened scalar potential disorder parameter 
$\tgu$ [defined by Eq.~(\ref{tguDef})]. The ratio of the other disorder strengths
$g_{A}$ and $g_{m}$ to $\tgu$ asymptotes toward zero as the flow begins to diverge.
For stronger values of the Coulomb interaction strength $\bar{w} \gtrsim 0.1$ (for the choice $\eta = 0.3$),
we observe a ``crossover'' in the flow direction. In the limit of large Coulomb strength $\bar{w} \gtrsim 1$, we 
find that the flow becomes dominated by the divergence of the SU(2) non-Abelian vector potential disorder 
parameter $g_{A}$; in this regime, the ratio of the other disorder strengths $\tgu$ and $g_{m}$ to $g_{A}$ 
asymptotes to zero as the flows leave the perturbatively accessible regime.

We demonstrate the picture described above with a selection of RG flow trajectory plots. 
Figs.~\ref{RGFlow1--Origin}--\ref{RGFlow2--guElev} depict two sets of flow trajectories obtained 
via numerical integration of Eqs.~(\ref{lnwFlowVSD})--(\ref{gmFlowVSD}) for different initial 
Coulomb interaction strengths $\bar{w}$. Figs.~\ref{RGFlow1--Origin} and \ref{RGFlow2--gu} depict
projections of these flows in the 3D coupling constant subspace $(\tgu, g_{A}, \bar{w})$. 
Figs.~\ref{RGFlow1--OriginElev} and \ref{RGFlow2--guElev} respectively exhibit 2D projections of the same flows 
represented in Figs.~\ref{RGFlow1--Origin} and \ref{RGFlow2--gu} in the disorder plane 
$(\tgu, g_{A})$. We choose \emph{not} to exhibit the RG evolution of the random mass parameter $g_{m}$ 
[Eqs.~(\ref{DisDef}), (\ref{DisVar}), and (\ref{gmFlowVSD})] in these figures, because we find
that $g_{m}$ always plays a subleading role relative to either the scalar $\tgu$ or non-Abelian vector 
$g_{A}$ potential disorder strengths. 

In each of Figs.~\ref{RGFlow1--Origin}--\ref{RGFlow2--guElev},
all trajectories share a given set of initial disorder strengths (described in the figure captions).
Different trajectories, distinguished by lowercase Latin letters, correspond
to different initial Coulomb interaction strengths $\bar{w}(0)$. Trajectories with successive labels 
(a), (b), etc.\ carry successively larger initial values of $\bar{w}$; numerical values are
stated in the captions of Figs.~\ref{RGFlow1--Origin} and \ref{RGFlow2--gu}. Although very weak
initial Coulomb interaction strengths $\bar{w} \lesssim 0.1$ lead to trajectories that 
are dominated by large scalar potential disorder fluctuations $\tgu \rightarrow \infty$, stronger
initial interaction strengths bend the trajectories away from the $\tgu$-$\bar{w}$ plane, toward
the particle-hole symmetric $g_{A}$-$\bar{w}$ plane. [The \emph{effective} particle-hole symmetry is 
defined by the $\textrm{PH}^{*}$ transformation given by Eq.~(\ref{PHstar})]. 

These figures encapsulate our primary result: non-Abelian vector potential disorder 
dominates the flow toward strong coupling ($g_{A},\bar{w} \rightarrow \infty$) for moderate 
to strong Coulomb interaction strengths. As stated above, this result is essentially 
independent of our expansion parameter $\eta$, defined by Eq.~(\ref{etaDef2}), for
our one-loop flow equations (\ref{FlowEqs}) and (\ref{FlowEqsVSD}), obtained to lowest nontrivial order
in $1/N$ and in the disorder strengths $\tgu$, $g_{A}$, $g_{m}$. Physically, strong scalar 
potential disorder fluctuations would favor the accumulation of electrons and holes in 
spatially segregated ``puddles,'' locally violating charge neutrality. The advent of a 
second type of disorder that manifestly preserves a type of particle-hole symmetry
[Eq.~(\ref{PHstar})] allows the system to flow toward a strongly disordered, interacting 
regime, while everywhere preserving electroneutrality. Equivalently, Coulomb interparticle 
interactions lead to screening effects, which parametrically curtail the scalar potential 
disorder fluctuations; the non-Abelian vector potential is not affected by the interactions 
because it represents charge-neutral randomness (such as the vector potential component of 
``ripples''
).

To close this section, we direct the reader's attention to Fig.~\ref{PhaseDiag}.
In this figure, we exhibit a putative phase diagram for graphene, projected into
the 3D $(\tgu,g_{A},\bar{w})$ disorder-interaction coupling strength space, based
upon our work here, as well as previous 
results\cite{Finkelstein,CastellaniDiCastroLeeMaSorellaTabet,BK,NersesyanTsvelikWenger,OstrovskyGornyiMirlin1}
regarding the zero temperature ground state physics of 2D disordered and/or interacting electronic systems. 
The most important features of this figure are the points labeled (a)--(c) and the ``globes'' 
labeled (d)--(f). Points (a)--(c) label known phases of the graphene model in the \emph{absence} of interactions: 
(a) is the non-disordered, non-interacting Dirac theory. (b) denotes the critical, delocalized 
phase that occurs at strong non-Abelian vector potential (``chiral'') disorder, 
described analytically by the class CI conformal field theory.\cite{NersesyanTsvelikWenger,CILogCFT} As discussed
in Secs.~\ref{ClassCIandBDI} and \ref{ClassCIRG}, the class CI fixed point 
is expected to possess a non-zero conductance at zero temperature that is 
independent of the disorder strength.\cite{OstrovskyGornyiMirlin1}
Finally, (c) denotes the Anderson insulating phase for the orthogonal normal metal
quantum disorder class. Globe (d) represents a ``theory arena'' corresponding to the symplectic 
(spin-orbit) normal metal class for non-interacting electrons, 
described by the symplectic non-linear
sigma model 
(possibly modified by a topological 
term).\cite{OstrovskyGornyiMirlin2,BardarsonTworzydloBrouwerBeenakker,NomuraKoshinoRyu,RyuMudryObuseFurusaki}
The symplectic class may possess both metallic and insulating states in 2D.\cite{LeeRamakrishnan} 

Globes (e) and (f) respectively describe the symplectic and orthogonal normal metal quantum disorder
classes, augmented with long-range Coulomb interparticle 
interactions.\cite{Finkelstein,CastellaniDiCastroLeeMaSorellaTabet,BK,CastellaniDiCastroForgacsSorella,KotliarSorella}
These globes describe physics on temperature scales less than the elastic scattering rate, i.e.\ 
$T \lesssim 1/\tau_{el}$, with $\tau_{el}$ the elastic transport lifetime due to impurity scattering.
The low-energy field theory description takes the form of a modified non-linear
sigma model that incorporates electron-electron interactions.\cite{Finkelstein,BK} 
The Drude conductivity computed in the regime $T \sim \tau_{el}$, discussed in the next section,   
enters through the bare diffusion constant of the sigma model.

Arrows in Fig.~\ref{PhaseDiag} schematically denote RG flows whose topology has either been derived in this 
paper, previous work,\cite{BK} or may be inferred by general principles.
The dashed line residing in the $\tgu$--$\bar{w}$ plane represents the repulsive fixed line described in 
Sec.~\ref{ClassAIIRG}; the section of the $\tgu$--$\bar{w}$ plane between this line and the Coulomb
interaction ($\bar{w}$) axis drains into the clean, non-interacting Dirac fixed point (a). This 
section is, however, highly unstable in the other disorder (i.e.\ $g_{A}$) directions.
The reader should imagine ``superimposing'' the flow trajectories depicted in Figs.~\ref{RGFlow1--Origin} 
and \ref{RGFlow2--gu} upon this figure.


\section{Physical Results and Discussion\label{Physics}}

In this section, we articulate a number of predictions for the scaling behavior of the dc electrical
conductivity, as well as the thermal transport, for graphene in its Drude/Boltzmann transport regime.
The predictions are extracted from the large-$N$, weak disorder renormalization group (RG) analysis performed 
above. 
We presuppose here that the temperature is sufficiently high so that quantum interference corrections 
to the conductivity due to electronic diffusion (weak localization and Altshuler-Aronov--type interaction effects) 
may be ignored.\cite{LeeRamakrishnan,AltshulerAronov,AleinerAltshulerGershenson}

We restore Planck's constant via $h = 2\pi$ throughout this section.

\subsection{Disorder vs.\ interaction limited transport coefficients}

The kinetic transport coefficients in graphene are explicit functions of the effective disorder and
interaction strengths, depending upon both the elastic transport lifetime $\tau_{el}$ due to impurity scattering,
and upon the inelastic transport lifetime $\tau_{in}$ due to electron-electron collisions. 
Estimates for these are
\begin{subequations}
\begin{align}\label{LifetimeEst}
	\frac{\hbar}{\tau_{el}} &\propto \bGt \, \textrm{max}(|\mu|,T),
	\\
	\frac{\hbar}{\tau_{in}} &\propto \frac{1}{N}\, \textrm{min}(\bar{w}^2,1) \, \textrm{min}(T,\frac{T^2}{|\mu|}),
\end{align}
\end{subequations}
where $\mu$ is the chemical potential, $T$ is the temperature, and $\bGt$ is a certain combination
of the disorder parameters in Eq.~(\ref{DisVar})--see Eqs.~(\ref{DrudeEl}) and 
(\ref{GTransportDef}), below. [Recall that $\bar{w} \propto r_{s}$, Eq.~(\ref{wbarrs}).]

We consider the two limits with large and small $\tau_{el}/\tau_{in}$ in turn:

\subsubsection{$\tau_{el} > \tau_{in}$: interaction-limited}

Real, inelastic electron-electron collisions play an
important role in determining the kinetic coefficients for $\tau_{el} > \tau_{in}$. 
In particular, at exactly zero doping $\mu = 0$, the composite electron-hole fluid
is electrically neutral. Here, electron-hole collisions are enough to
set a finite dc conductivity $\sigma_{\mathsf{dc}}$, which takes the form 
\begin{equation}\label{DrudeInel}
	\sigma_{dc} \propto \frac{N^2 e^2}{h} \frac{1}{\textrm{min}(\bar{w}^2,1)}. 
\end{equation}
For Coulomb interaction strengths $\bar{w} \gtrsim 1$, Eq.~(\ref{DrudeInel}) predicts a ``minimum
metallic conductivity'' that is independent of both the interaction and disorder
strengths (to a first approximation). The doping dependence of the conductivity
is limited by the impurity scattering.

Violations of Mott's formula [Eq.~(\ref{ThermoPWRMott})]
for the thermopower and of the Wiedemann-Franz law for the thermal conductivity
are expected in the interaction-limited regime. Quantitative results based on the 
relativistic hydrodynamics description of the electron-hole plasma in graphene
will appear elsewhere.\cite{Graphene2--Hydrodynamics}

\subsubsection{$\tau_{el} < \tau_{in}$: disorder-limited}

When $\tau_{in}$ due to electron-electron collisions exceeds $\tau_{el}$ due to 
the impurities, the situation is simpler. To the first approximation, we can 
neglect contributions to the kinetic coefficients due to inelastic processes.
The Drude conductivity in the ladder approximation then takes the form
\begin{equation}\label{DrudeEl}
	\sigma_{\mathsf{dc}} = \frac{N e^{2}}{h} E_{F} \, \tau_{el},
\end{equation}
where\cite{Graphene2--Hydrodynamics}
\begin{subequations}\label{GTransportDef}
\begin{align}
	E_{F} \, \tau_{el} &= \frac{1}{\pi \bGt},
	\\
	\bGt &\equiv \tgu + 8 g_{A} + 4 g_{A3} + 6 g_{m} + 3 g_{v}.
\end{align}
\end{subequations}
Note that it is the \emph{screened} scalar potential disorder parameter $\tgu$ 
[Eq.~(\ref{tguDef})] that appears in Eq.~(\ref{GTransportDef}).
The temperature or chemical potential-dependence of the conductivity follows
from combining the scaling predictions of Sec.~\ref{Results} with Eq.~(\ref{DrudeEl});
Mott's formula and Wiedemann-Franz apply for thermotransport. 
In the remainder of this section, we will use this strategy to identify three
scaling regimes for the transport coefficients in the disorder-limited case.

\subsection{Scaling predictions for graphene transport}

The essential message is as follows: if the disorder distribution is sufficiently weak and short-range
correlated, then the RG generically predicts corrections to the conductivity that are logarithmic in temperature.
In particular, we have identified three scaling regimes 
which may be observed in (future) graphene experiments: 
(1) a ``QED'' regime, wherein the conductivity \emph{increases} with decreasing temperature, as well as (2) 
intermediate and (3) ``QCD'' regimes, characterized by a conductivity that decreases monotonically with decreasing
temperature. Here, the descriptors ``QED'' and ``QCD'' refer to formal analogies between the disordered, interacting
graphene theory in particular scaling regimes, and the theories of high energy quantum electro- and chromodynamics,
respectively; we do \emph{not} imply a mapping between experimental phenomena in high energy particle and graphene 
physics.  

In this subsection, we set $N = 4$, appropriate to real graphene. To spare the reader from technical details, 
the derivation of the formulae presented below has been relegated to Appendix~\ref{APP-Physics}.
A sketch of our results is pictured in Fig.~\ref{ConductivityScaling}.

\subsubsection{``QED'' regime: Scalar potential disorder and flow toward weak coupling\label{IntWeak}}

The bare Coulomb interaction strength $r_{s} = e^{2}/\epsilon \hbar v_{F}$ is of order unity in 
the previous substrate-supported experiments,\cite{Geim1,Geim2,Kim,Kim2} and should exceed two
for suspended films.\cite{MeyerGeim} As discussed in Secs.~\ref{Intro} and \ref{Results}, an 
$r_{s} \sim 1$ represents a relatively strong coupling regime for the massless Dirac electrons 
in undoped graphene. If the disorder effects are weak relative to the Coulomb interaction
strength, then the RG predicts the existence of a ``QED'' scaling regime, wherein disorder and 
electron correlation effects grow ever \emph{weaker} on larger length or lower energy scales.
This scaling toward weak coupling manifests itself as a logarithmic-squared increase in the
dc conductivity with decreasing temperature [Eq.~(\ref{DrudeIntWeakCoupling}), below]. A further 
requirement to reach the ``QED'' scaling regime is that scalar potential disorder must provide 
the dominant scattering mechanism for the Dirac electrons.
(Consult Secs.~\ref{ClassAII}--\ref{ClassD}, \ref{ClassAIIRG}, and the last paragraph of 
Sec.~\ref{GrapheneResults}.)

In this ``QED'' regime, we find that (see Appendix~\ref{APP-Physics} for details)
\begin{equation}\label{DrudeIntWeakCoupling}
	\sigma_{\mathsf{dc}}(T) \sim \frac{4 e^{2}}{h \pi} 
	\frac{r_{s}^{2}}{16 \tgu} \ln^{2}\left(\frac{T_{R}}{T}\right),
\end{equation}
where $T_{R} \gg T$ is an arbitrary reference temperature,
and $\tgu \ll r_{s} \sim 1$ is the microscopic strength of the (screened) scalar potential disorder 
[Eqs.~(\ref{DisDef})--(\ref{DisVar}), (\ref{tguDef})]. 

\begin{figure}
\includegraphics[width=0.35\textwidth]{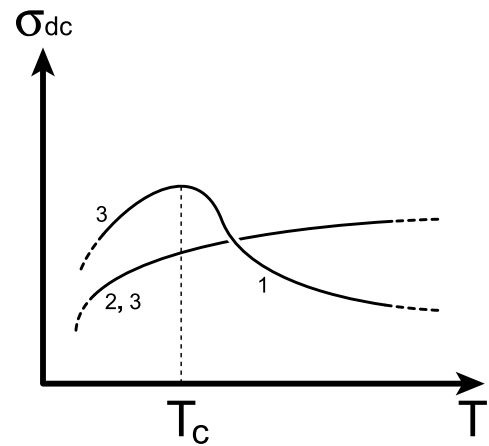}
\caption{Sketch of possible logarithmic scaling regimes of the dc conductivity 
$\sigma_{\mathsf{dc}}$ vs.\ temperature in undoped graphene. The labels $(1)$--$(3)$ identify 
the qualitative scaling behaviors pictured here with the three regimes respectively discussed in subsections 
\ref{IntWeak}--\ref{AsymStrong} in the text. Non-monotonic behavior, i.e.\ the peak 
separating the curve segments labeled $(1)$ and $(3)$ in the figure, can occur if 
a sample microscopically dominated by strong Coulomb interactions and weak scalar potential disorder
(``QED'' regime), distinguished by a conductivity that increases with decreasing temperature 
[Eq.~(\ref{DrudeIntWeakCoupling})], crosses over at lower temperatures to the strong coupling 
(``QCD'') regime dominated by non-Abelian vector potential disorder, characterized by a 
decreasing conductivity [Eq.~(\ref{DrudeAsymStrongCouplingSMALLgA})]. An intermediate scaling regime 
$(2)$ is also possible, characterized by a monotonically decreasing 
conductivity [Eq.~(\ref{DrudeIntStrongCoupling})]. 
These predictions require that $\tau_{el} < \tau_{in}$; see text.
\label{ConductivityScaling}}
\end{figure}

The behavior described by Eq.~(\ref{DrudeIntWeakCoupling}) cannot persist indefinitely,
because the presence of other disorder parameters will eventually induce a crossover to the
``QCD'' regime described below in subsection \ref{AsymStrong}. This crossover would be distinguished 
by a \emph{non-monotonicity} of the conductivity, as shown in Fig.~\ref{ConductivityScaling}. 
We do not provide here an estimate for the non-universal crossover temperature $T_{c}$, which is difficult 
to extract quantitatively because it depends upon the details of the full microscopic
disorder distribution.\cite{footnote-e}

We also discuss thermotransport. We neglect interaction contributions or corrections to the 
energy current of the Dirac electrons; as a result, we find that the thermal conductivity 
is slaved to Eq.~(\ref{DrudeIntWeakCoupling}) via the Wiedemann-Franz law. 
Within this same approximation, the thermopower $S$ is given by Mott's formula\cite{Mahan} ($k_{B} = 1$):
\begin{equation}\label{ThermoPWRMott}
	S \sim -\frac{\pi^{2}}{3}\frac{T}{e}
		\frac{d \ln \sigma_{\mathsf{dc}}(\epsilon)}{d\epsilon}.
\end{equation}
In this expression, $\sigma_{\mathsf{dc}}(\epsilon)$ is the scaling form of the conductivity,
with temperature replaced by $\epsilon$ (i.e.\ the chemical potential).
Using Eq.~(\ref{DrudeIntWeakCoupling}), we obtain
\begin{equation}\label{ThermoPWRIntWeakCoupling}
	S \sim \frac{\pi^{2}}{3}\frac{T}{e} \frac{2}{\mu \ln(\mu_{R}/\mu)},
\end{equation}
for $|\mu| > T$, with $\mu_{R} > \mu$.

\subsubsection{Intermediate regime: Scalar potential disorder and flow toward strong coupling\label{IntStrong}}

We next consider the case where the strength of the Coulomb interaction $r_{s}$ is made 
weaker than that of the screened scalar potential disorder fluctuations [$\tgu \gtrsim \bar{w} \sim r_{s}$]. 
We also require that scalar potential fluctuations provide the dominant source of scattering
for the Dirac electrons, while not becoming so strong so as to invalidate our weak disorder
assumption ($\tgu \lesssim 1$). Such a scenario might prove difficult to realize experimentally,
but could be achieved in principle by screening the long-ranged Coulomb interactions with an 
external gate (brought into very close proximity to the graphene layer), or by coating graphene 
with a high-k dielectric material.

The scaling behavior of the conductivity in this regime is given by
\begin{equation}\label{DrudeIntStrongCoupling}
	\sigma_{\mathsf{dc}}(T) \sim \frac{4 e^{2}}{h \pi}
	\left[
	\frac{1}{\tgu} - 2 \ln \left(\frac{T_{R}}{T}\right)
	\right],
\end{equation}
for $T_{R} > T$ (Appendix~\ref{APP-Physics}). 

The thermopower at non-zero chemical potential $|\mu| > T$ is given by
\begin{equation}\label{ThermoPWRIntStrongCoupling}
	S \sim -\frac{\pi^{2}}{3}\frac{T}{e} \frac{2 \tgu}{\mu}.
\end{equation}

\subsubsection{``QCD'' regime: Non-Abelian vector potential disorder and asymptotic flow 
toward strong coupling\label{AsymStrong}}

Non-abelian vector potential disorder provides another potentially important scattering
mechanism in graphene.\cite{MorozovNovoselov,MeyerGeim,MorpurgoGuinea} This disorder type
arises in the context of elastic deformations or ``ripples''  
and in the context of topological lattice defects.\cite{TopologicalDisorder}
(See also Secs.~\ref{ClassCIandBDI} and \ref{ClassCIRG}.)

Surprisingly, we have found in this paper that quenched vector potential
disorder generically emerges as the dominant scattering mechanism at long length scales and low energies
for Dirac electrons with a bare Coulomb interaction parameter $r_{s} \gtrsim 0.1$. [See
Sec.~\ref{GrapheneResults} for details.] In this regime, the graphene theory becomes formally analogous
to $3+1$-D quantum chromodynamics (QCD), in that the system becomes dominated by strong non-Abelian 
vector potential disorder and electron correlation effects (due to the Coulomb interactions).
The essential difference between graphene and real QCD is that in the former (latter) case, it is a 
static, time-independent (dynamic, gluon-mediated) non-Abelian gauge potential that drives the 
system toward strong coupling at low energies.  

Since $r_{s} \gtrsim 1$ in real graphene, we generically expect to reach the ``QCD'' regime
at some finite temperature scale. We obtain the following formula for the 
asymptotic scaling behavior of the conductivity (Appendix~\ref{APP-Physics}):
\begin{equation}\label{DrudeAsymStrongCouplingSMALLgA}
	\sigma_{\mathsf{dc}}(T) \sim \frac{4 e^{2}}{h \pi}
	\left[
	\frac{1}{12 g_{A}} - \frac{2 \pi^{2}}{3(\pi^{2} - 2)} \ln \left(\frac{T_{R}}{T}\right)
	\right],
\end{equation}
where $g_{A}$ is the (renormalized) strength of the non-Abelian vector potential disorder,
and $T_{R} > T$. Eq.~(\ref{DrudeAsymStrongCouplingSMALLgA}) is not expected to hold down to arbitrarily
low temperatures. As we examine lower and lower energy scales, the effective vector potential disorder 
strength $g_{A}(T)$ grows ever larger (Sec.~\ref{GrapheneResults}); as a result, the scaling behavior
of Eq.~(\ref{DrudeAsymStrongCouplingSMALLgA}) will mutate as $g_{A}$ evolves. [A more general scaling
formula in the ``QCD'' regime is given by Eq.~(\ref{APP-DrudeAsymStrongCoupling}) in Appendix~\ref{APP-Physics}, 
which reduces to Eq.~(\ref{DrudeAsymStrongCouplingSMALLgA}) in the limit $g_{A} \ll 1$.]

The thermopower at finite chemical potential $|\mu| > T$ corresponding to
Eq.~(\ref{DrudeAsymStrongCouplingSMALLgA}) is given by
\begin{equation}\label{ThermoPWRAsymStrongCouplingSMALLgA}
	S \sim -\frac{\pi^{2}}{3}\frac{T}{e} \frac{8 g_{A}}{\mu} \frac{\pi^{2}}{\pi^{2} - 2}.
\end{equation}

Finally, we reiterate the following point: we have assumed that the inelastic transport lifetime due 
to electron-electron collisions, $\tau_{in}$, exceeds the elastic lifetime $\tau_{el}$ due to the disorder. 
In graphene, the kinetic coefficients are sensitive to the value of $\tau_{in}/\tau_{el}$; we will
discuss the implications of the above three scaling regimes in the limit $\tau_{in}/\tau_{el} > 1$
in a subsequent publication.\cite{Graphene2--Hydrodynamics}


\begin{acknowledgments}

We would like to thank Andreas Ludwig for helpful discussions at the beginning of this work.
One of us (MSF) was supported primarily by the Nanoscale Science and Engineering Initiative of the
National Science Foundation under NSF Award Number CHE-06-41523, and by the New York State Office 
of Science, Technology, and Academic Research (NYSTAR).

\end{acknowledgments}


\appendix


\section{Computation of the electronic self-energy to $\ord{\frac{1}{N}}$.\label{APP-PsiSLFNRG}}

In this appendix, we derive the amplitude given by Eq.~(\ref{D7.1}) for the diagram $\mathsf{D}_{5.1}$,
pictured in Fig.~\ref{FigD7}. $\mathsf{D}_{5.1}$ is the contribution to the electronic self-energy due 
to the Coulomb interactions at order $1/N$. Using the Feynman rules from Sec.~\ref{LargeN}, one has
\begin{align}
	\mathsf{D}_{5.1} =& \left(i \sqrt{\frac{w}{N}}\right)^{2}
	\begin{aligned}[t]
	\int 
	\frac{d\omega \, \dvex{l}}{(2\pi)^{3}} \,
	&
	\frac{i (\omega_{n}+\omega) + v_{F}\sigmah\cdot(\bm{\mathrm{k}}+\bm{\mathrm{l}})}
		{(\omega_{n}+\omega)^{2} + v_{F}^2 (\bm{\mathrm{k}}+\bm{\mathrm{l}})^{2}}
	\\
	&\times
	\frac{1}{|\bm{\mathrm{l}}|}
	\frac{\sqrt{v_{F}^2 l^{2}+\omega^{2}}}{\sqrt{v_{F}^2 l^{2} + \omega^{2}}+ \bar{w} v_{F} |\bm{\mathrm{l}}|}.
	\end{aligned}
\end{align}
Expanding in terms of the external frequency and momentum, we obtain
\begin{equation}\label{D7.1--II}
	\mathsf{D}_{5.1} \sim
	\frac{-w}{v_{F} N}\left[i \omega_{n} \, I_{1} + v_{F} \sigmah\cdot\vex{k} \, I_{2}\right],
\end{equation}
where
\begin{subequations}
\begin{align}
	I_{1} &= \int \frac{d\omega \, \dvex{l}}{(2\pi)^{3}} \,
	\frac{1}{|\bm{\mathrm{l}}|}
	\frac{\sqrt{l^{2}+\omega^{2}}}{\sqrt{l^{2}+\omega^{2}}+ \bar{w} |\bm{\mathrm{l}}|}
	\frac{\vex{l}^{2} - \omega^{2}}{\left(\omega^{2} + \vex{l}^{2}\right)^{2}},
	\label{I1Def}\\
	I_{2} &= \int \frac{d\omega \, \dvex{l}}{(2\pi)^{3}} \,
	\frac{1}{|\bm{\mathrm{l}}|}
	\frac{\sqrt{l^{2}+\omega^{2}}}{\sqrt{l^{2}+\omega^{2}}+ \bar{w} |\bm{\mathrm{l}}|}
	\frac{\omega^{2}}{\left(\omega^{2} + \vex{l}^{2}\right)^{2}},
	\label{I2Def}\\
\intertext{and where it is useful to define}
	I_{3} &\equiv I_{1} + I_{2}
	\nonumber\\
	&= \int \frac{d\omega \, \dvex{l}}{(2\pi)^{3}} \,
	\frac{1}{|\bm{\mathrm{l}}|}
	\frac{\sqrt{l^{2}+\omega^{2}}}{\sqrt{l^{2}+\omega^{2}}+ \bar{w} |\bm{\mathrm{l}}|}
	\frac{\vex{l}^{2}}{\left(\omega^{2} + \vex{l}^{2}\right)^{2}}.
	\label{I3Def}
\end{align}
\end{subequations}
After rescaling $\omega \rightarrow \omega |\vex{l}|$, the
momentum integrals are trivially done, so that
\begin{subequations}
\begin{align}
	I_{2} &= 
	\frac{\ln\Lambda}{2 \pi^{2}}
	\int_{0}^{\infty} dx
	\frac{\sqrt{1+x^{2}}}{\sqrt{1+x^{2}}+ \bar{w}}
	\frac{x^{2}}{\left(x^{2} + 1\right)^{2}}
	\\
	I_{3}
	&= 
	\frac{\ln\Lambda}{2 \pi^{2}}
	\int_{0}^{\infty} dx
	\frac{\sqrt{1+x^{2}}}{\sqrt{1+x^{2}}+ \bar{w}}
	\frac{1}{\left(x^{2} + 1\right)^{2}}
\end{align}
\end{subequations}
where we neglect ultaviolet-finite terms associated with the infrared region
of the momentum integration.
We then find that
\begin{subequations}\label{I1I2I3}
\begin{align}
	I_{1} &\equiv \frac{\ln\Lambda}{2 \pi} f_{1}(\bar{w}), \label{I1}\\
	I_{2} &\equiv \frac{\ln\Lambda}{2 \pi} f_{2}(\bar{w}), \label{I2}\\
	I_{3} &\equiv \frac{\ln\Lambda}{2 \pi} f_{3}(\bar{w}), \label{I3}
\end{align}
\end{subequations}
where the functions $f_{1}$--$f_{3}$ were defined by Eqs.~(\ref{f1f2Def}) and (\ref{f3Def}).
Using Eq.~(\ref{I1I2I3}) in Eq.~(\ref{D7.1--II}), we recover Eq.~(\ref{D7.1}).


\section{Fixed lines and linearized flow equations\label{APP-FixedLines}}

The one-loop, large-$N$ flow equations given by Eq.~(\ref{FlowEqs}) in Sec.~\ref{Results}
possess a number of critical fixed line structures, continuously connected to the non-disordered, 
non-interacting Dirac fixed point [Eq.~(\ref{FreeDiracFixedPoint})]. In this appendix, we enumerate
these structures, and determine their stability in the six-dimensional coupling strength space
[c.f.\ Eq.~(\ref{FlowEqs})]. Each fixed line resides in a plane formed between the Coulomb 
interaction axis $\bar{w}$ and a single, nonzero disorder parameter.

\subsubsection{Scalar potential disorder}

As discussed in Sec.~\ref{ClassAIIRG}, in the case of pure scalar potential disorder,
i.e.\ only $\tgu > 0$ among the five disorder parameters appearing in Eq.~(\ref{FlowEqs}), the 
large-$N$ graphene model exhibits a repulsive fixed line in the $\tgu$--$\bar{w}$ plane. 
From Eq.~(\ref{FlowAII}), the fixed line is parameterized by the conditions
\begin{subequations}\label{guFixedLine}
\begin{align}
	g_{A} &= g_{A3} = g_{m} = g_{v} = 0,
	\\
	\tgun &= \eta \bar{w}^{(0)} f_{3}(\bar{w}^{(0)}).
\end{align}
\end{subequations}
Since $\tgu$ and $\eta = 8/\pi N$ are expansion parameters for our large-$N$ theory, the \emph{entirety}
of the fixed line described by Eq.~(\ref{guFixedLine}) is perturbatively accessible, because
the function $\bar{w} f_{3}(\bar{w})$ asymptotes to a finite constant as $\bar{w} \rightarrow \infty$;
see Eq.~(\ref{f3Lim}) and Fig.~\ref{fPlot}.

Linearizing Eq.~(\ref{FlowEqs}) about the line in Eq.~(\ref{guFixedLine}), we obtain the following RG 
eigenvalues (as a function of $\bar{w}^{(0)}$):
\begin{align}\label{EigenValues--gu}
	\{
	&
	0, \eta \bar{w}^{(0)} \left[\frac{2 f_{3}(\bar{w}^{(0)})}{1+ \bar{w}^{(0)}} - \mathcal{F}_{3}(\bar{w}^{(0)})\right],
	\nonumber\\
	&
	-2\sqrt{2} \, \eta \, \bar{w}^{(0)} \, f_{3}(\bar{w}^{(0)}), 2\sqrt{2} \, \eta \, \bar{w}^{(0)} \, f_{3}(\bar{w}^{(0)})
	\},
\end{align}
where we have introduced the function
\begin{align}\label{F3Def}
	\mathcal{F}_{3}(\bar{w}) &\equiv \frac{d \phantom{\bar{w}}}{d \bar{w}} 
	\left[\bar{w} f_{3}(\bar{w}) \right]
	\nonumber\\
	&=
	\frac{1}{2\bw^{2}}
	\left\lgroup
	\begin{aligned}
	&\frac{2}{\pi}
	\frac{1}{1 - \bar{w}^{2}}
	\\
	&\times\left[
	-\bar{w} + \frac{2\bar{w}^{2} - 1}{\sqrt{1-\bar{w}^{2}}}\arccos(\bar{w})
	\right]
	+ 1 
	\end{aligned}
	\right\rgroup.
\end{align}

The first two eigenvalues listed in Eq.~(\ref{EigenValues--gu}) respectively characterize RG flow trajectories 
along and through the fixed line within the $\tgu-\bar{w}$ plane; the latter two eigenvalues listed in this 
equation characterize the RG flow in the transverse directions, and are each doubly degenerate.
The second and fourth eigenvalues listed in Eq.~(\ref{EigenValues--gu}) are positive for all finite $\bar{w}^{(0)}>0$; 
both the $\tgun(\bar{w}^{(0)})$ fixed line, as well as the portion of the $\tgu$--$\bar{w}$ plane
spanning the region between this fixed line and the $\bar{w}$-axis are therefore highly unstable.

\subsubsection{Abelian vector potential disorder}

Next, we consider the following fixed line in the $g_{A3}$--$\bar{w}$ plane:
\begin{subequations}\label{gAFixedLine}
\begin{align}
	\tgu &= g_{A} = g_{m} = g_{v} = 0,
	\\
	g_{A3}^{(0)} &= \frac{\eta}{2} \bar{w}^{(0)} f_{3}(\bar{w}^{(0)}).
\end{align}
\end{subequations}
Linearizing Eq.~(\ref{FlowEqs}) along this line, we obtain the eigenvalues
\begin{align}\label{EigenValues--gA3}
	\{
	&
	0, -\eta \, \bar{w}^{(0)} \, \mathcal{F}_{3}(\bar{w}^{(0)}), 2 \eta \bar{w}^{(0)} \, f_{3}(\bar{w}^{(0)}), 
	4 \eta \bar{w}^{(0)} \, f_{3}(\bar{w}^{(0)}),
	\nonumber\\
	&
	\frac{2 \eta \, \bar{w}^{(0)} \, f_{3}(\bar{w}^{(0)})\left[1 - \bar{w}^{(0)} \pm 
	\sqrt{2+\bar{w}^{(0)}(\bar{w}^{(0)}-2)}\right]}{1+\bar{w}^{(0)}}
	\}.
\end{align}
The first two eigenvalues listed in Eq.~(\ref{EigenValues--gA3}) respectively characterize RG flow 
trajectories along and through the fixed line within the $g_{A3}$--$\bar{w}$ plane, where it serves 
as a (locally) attractive, critical extension of the clean Dirac fixed point [located by 
Eq.~(\ref{FreeDiracFixedPoint})]. The latter four eigenvalues appearing in Eq.~(\ref{EigenValues--gA3}) 
characterize the RG flow in the transverse directions, and show that this fixed line is highly unstable.

\subsubsection{Staggered potential disorder}

Finally, we consider a fixed line in the $g_{v}$--$\bar{w}$ plane, parameterized by the
conditions
\begin{subequations}\label{gvFixedLine}
\begin{align}
	\tgu &= g_{A} = g_{A3} = g_{m} = 0,
	\\
	g_{v}^{(0)} &= \eta \bar{w}^{(0)} f_{3}(\bar{w}^{(0)}).
\end{align}
\end{subequations}
Linearizing Eq.~(\ref{FlowEqs}) yields the RG eigenvalues
\begin{align}\label{EigenValues--gv}
	\{
	&
	0, -\eta \, \bar{w}^{(0)} \left[2 f_{3}(\bar{w}^{(0)}) + \mathcal{F}_{3}(\bar{w}^{(0)})\right],
	\nonumber\\
	&
	2(1 \pm \sqrt{3}) \eta \, \bar{w}^{(0)} \, f_{3}(\bar{w}^{(0)}),
	\pm \frac{2\sqrt{2} \eta \bar{w}^{(0)} f_{3}(\bar{w}^{(0)})}{1+\bar{w}^{(0)}}
	\}.
\end{align}
The first two eigenvalues listed in this equation respectively characterize RG flow trajectories along 
and through the fixed line within the $g_{v}$--$\bar{w}$ plane; the latter four characterize the RG flow in 
the transverse directions. We observe that the $g_{v}^{(0)}(\bar{w}^{(0)})$ critical fixed line is 
also unstable for all finite $\bar{w}^{(0)} > 0$.


\section{Derivation of physical results\label{APP-Physics}}

\begin{figure}
\includegraphics[width=0.4\textwidth]{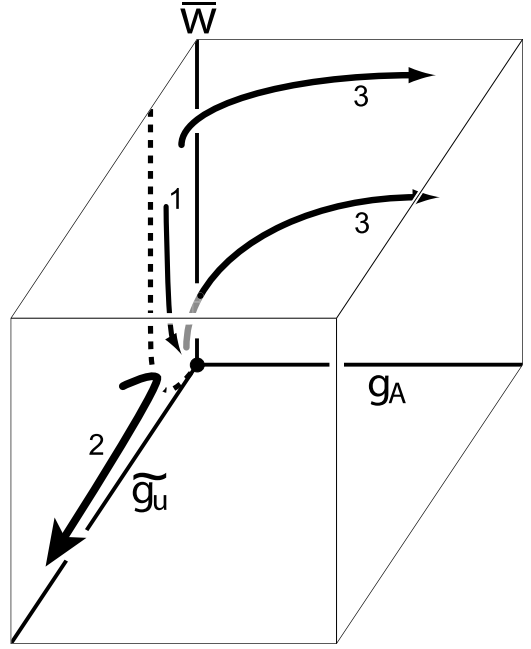}
\caption{Schematic RG flow diagram, projected into the 3D coupling constant space formed from
the screened scalar potential disorder strength ($\tgu$), the non-Abelian vector potential disorder
strength ($g_{A}$), and Coulomb interaction strength ($\bar{w} = \pi r_{s}/2$) [Eq.~(\ref{wbarrs})].
This diagram should be compared to the flow plots, obtained via numerical integration
of Eq.~(\ref{FlowEqsVSD}), depicted in Figs.~\ref{RGFlow1--Origin} and \ref{RGFlow2--gu} 
of Sec.~\ref{GrapheneResults}. In this figure, we have sketched characteristic flows, designated
by the labels $(1)$--$(3)$, corresponding to the ``QED,'' intermediate, and ``QCD'' scaling regimes
discussed in Sec.~\ref{Physics}.
The dashed curve represents the unstable fixed line discussed below Eq.~(\ref{FlowAII}) in 
Sec.~\ref{ClassAIIRG}.
\label{ConductivityScalingRG}}
\end{figure}

In this appendix, we derive the scaling predictions for the dc conductivity stated in 
Eqs.~(\ref{DrudeIntWeakCoupling}), (\ref{DrudeIntStrongCoupling}), and 
(\ref{DrudeAsymStrongCouplingSMALLgA}) of Sec.~\ref{Physics}. These predictions arise in the 
``QED,'' intermediate, and ``QCD'' scaling regimes respectively, as defined in that section.

\subsection{``QED'' regime: Scalar potential disorder and flow toward weak coupling\label{APP-IntWeak}}

The ``QED'' regime occurs when the microscopic initial conditions are such that the (screened) scalar potential 
disorder fluctuations are much stronger than that of all other disorder types: 
$\tgu \gg \mathcal{G}_{\mu}^{\nu}$ [see Eqs.~(\ref{DisDef})--(\ref{DisVar}), (\ref{tguDef}), and (\ref{DisMetric})]. 
Over some intermediate window of scaling, we would expect the RG to be dominated by $\tgu$ and the 
Coulomb interaction strength $\bar{w}$; to that end, we set all other disorder
parameters to zero, and integrate the RG flow described by Eq.~(\ref{FlowAII}).

As discussed in Sec.~\ref{ClassAIIRG}, these flow equations possess a repulsive fixed line,
pictured as the dashed curve in Figs.~\ref{PhaseDiag} and \ref{ConductivityScalingRG}. At the end of 
Sec.~\ref{GrapheneResults}, we noted that this line divides the $\tgu$--$\bar{w}$ plane into two 
portions; the portion joining the fixed line to the $\bar{w}$-axis flows back to the clean, 
non-interacting Dirac fixed point, while its complement flows toward strong disorder and interaction 
coupling. (The portion that flows back to the clean Dirac fixed point is destabilized by the other disorder
parameters for arbitrarily weak but non-zero disorder strength, as shown in Appendix~\ref{APP-FixedLines}.)
The RG evolves toward weak (strong) coupling at longer length and lower energy scales if the 
``microscopic'' disorder and interaction strengths
$\tgu$ and $\bar{w}$ satisfy $\tgu < \eta \bar{w} f_{3}(\bar{w})$ 
[$\tgu > \eta \bar{w} f_{3}(\bar{w})$].

In the case of a flow toward weak coupling (``QED'' regime), the Coulomb interaction $\bar{w}$ rapidly 
decreases toward zero; we therefore take $\bar{w} \ll 1$ without loss of generality.
One obtains the scaling behavior [using RG Eqs.~(\ref{lnvFFlow}), 
(\ref{APP-TempScaling}), and (\ref{FlowAII}), and Eqs.~(\ref{DrudeEl}) and (\ref{GTransportDef}) for $\sigma_{\mathsf{dc}}$]
\begin{equation}\label{APP-DrudeIntWeakCoupling}
	\sigma_{\mathsf{dc}}(T) \sim \frac{N e^{2}}{2 \pi^{2}} 
	\frac{(\eta \, \bar{w})^{2}}{16 g_{u}} \ln^{2}\left(\frac{T_{R}}{T}\right),
\end{equation}
where $\eta = 8/\pi N$ and $T_{R} \gg T$ is an arbitrary reference temperature. Setting $N = 4$,
$g_{u} \propto \tgu$, and 
replacing $\bar{w} \rightarrow \pi r_{s}/2$ [Eq.~(\ref{wbarrs})],
we obtain Eq.~(\ref{DrudeIntWeakCoupling}) in Sec.~\ref{Physics}.

The evolution towards weak coupling characteristic of the ``QED'' regime is represented via the
schematic RG flow labeled $(1)$ in Fig.~\ref{ConductivityScalingRG}. Although this flow approaches 
the clean, non-interacting Dirac fixed point [located by Eq.~(\ref{FreeDiracFixedPoint}) in
Sec.~\ref{GrapheneResults}], it is ultimately deflected toward the strong-coupling, ``QCD'' regime, labeled (3) 
in Fig.~\ref{ConductivityScalingRG}, for any non-zero, but arbitrarily small
random vector potential ($g_{A}$) or random mass ($g_{m}$) disorder fluctuations.

\subsection{Intermediate regime: Scalar potential disorder and flow toward strong coupling\label{APP-IntStrong}}

We next consider the case where again $\tgu \gg \mathcal{G}_{\mu}^{\nu}$, but now 
$\tgu > \eta \bar{w} f_{3}(\bar{w})$, so that the RG flows toward strong coupling
[Eq.~(\ref{FlowAII})]. We further restrict $\bar{w} \ll 1$, since larger values of
$\bar{w}$ induce a crossover to the ``QCD'' regime dominated by the vector potential disorder,
described in the next subsection.

With these assumptions, the intermediate scaling behavior of the conductivity is 
given by
\begin{equation}\label{APP-DrudeIntStrongCoupling}
	\sigma_{\mathsf{dc}}(T) \sim \frac{N e^{2}}{2 \pi^{2}}
	\left[
	\frac{1}{\tgu} - 2 \ln \left(\frac{T_{R}}{T}\right)
	\right],
\end{equation}
for $T_{R} > T$. We recover Eq.~(\ref{DrudeIntStrongCoupling}) for $N = 4$. 

The evolution of the RG in this intermediate regime is represented by the curve
labeled (2) in Fig.~\ref{ConductivityScalingRG}.

\subsection{``QCD'' regime: Non-Abelian vector potential disorder and asymptotic 
flow toward strong coupling\label{APP-AsymStrong}}

Finally, we consider the flow toward strong coupling for moderate to strong Coulomb
interactions. This is the ``QCD'' regime dominated by the non-Abelian vector potential 
disorder parameters $g_{A}$ and $g_{A3}$. As discussed in Sec.~\ref{GrapheneResults},
valley space SU(2) rotational symmetry is restored on average as the RG scales toward
strong coupling;  we may therefore take $g_{A} \sim g_{A3}$. 

Neglecting all disorder parameters except $g_{A}$, we integrate Eqs.~(\ref{lnvFFlowVSD}),
(\ref{lnwFlowVSD}), and (\ref{gAFlowVSD}), and combine the results with 
Eqs.~(\ref{APP-TempScaling}), (\ref{DrudeEl}), and (\ref{GTransportDef}).
For $\bar{w} \gtrsim 1$, we obtain the following formula for the asymptotic scaling
behavior of the conductivity:
\begin{equation}\label{APP-DrudeAsymStrongCoupling}
	\sigma_{\mathsf{dc}}(T) \sim \frac{N e^{2}}{2 \pi^{2}}
	\frac{1}{12 g_{A} \phi} 
	W\left[
	\phi \, e^{\phi}
	\left(
	\frac{T}{T_{R}}
	\right)^{4/3}
	\right],
\end{equation}
for $T_{R} > T$. In this equation, 
\begin{equation}
	\phi \equiv \frac{1}{6 g_{A}}\left(1 - \frac{\eta}{\pi}\right),
\end{equation}
and $W(z)$ is Lambert's function, i.e.\ solves the equation $W e^{W} = z$.
For $g_{A} \ll 1$, Eq.~(\ref{APP-DrudeAsymStrongCoupling}) reduces to 
\begin{equation}\label{APP-DrudeAsymStrongCouplingSMALLgA}
	\sigma_{\mathsf{dc}}(T) \sim \frac{N e^{2}}{2 \pi^{2}}
	\left[
	\frac{1}{12 g_{A}} - \frac{2 \pi}{3(\pi - \eta)} \ln \left(\frac{T_{R}}{T}\right)
	\right],
\end{equation}
which is the same as Eq.~(\ref{DrudeAsymStrongCouplingSMALLgA}) for $N = 4$ ($\eta = 2/\pi$).

The curves labeled (3) in Fig.~\ref{ConductivityScalingRG} schematically indicate the
flow toward strong coupling associated with the scaling prediction given by 
Eq.~(\ref{APP-DrudeAsymStrongCouplingSMALLgA}) in this ``QCD'' regime; see also 
Figs.~\ref{RGFlow1--Origin}--\ref{RGFlow2--guElev} in Sec.~\ref{GrapheneResults}.



\begin{thebibliography}{99}
\bibitem{footnote-a} We employ the conventional basis for all Pauli matrices.
\bibitem{footnote-c}The key ingredient leading to symplectic, rather than orthogonal symmetry, is the factorization of the symmetric 
	matrix $\sih{2} \kah{2}$ into antisymmetric $\sih{2}$ and $\kah{2}$ parts. The absence of valley-space symmetry 
	breaking [due to the absence of disorder types in Eq.~(\ref{DisDef}) other than $u(\vex{r})$] converts 
	Eq.~(\ref{TRI}) into (\ref{TRIstar}). In terms of the single-particle Hamiltonian $\hat{h}$, imposing Eq.~(\ref{TRI}) or 
	(\ref{TRIstar}) respectively removes an orthogonal or symplectic subalgebra from the parent unitary Lie algebra,
	placing $\hat{h}$ into the orthogonal or symplectic Wigner-Dyson class, respectively.
\bibitem{footnote-e}The crossover between the ``QED'' and ``QCD'' regimes described in this section occurs through
	the RG flow equations [Eqs.~(\ref{FlowEqs}) and (\ref{FlowEqsVSD})] in a way that involves
	all of the disorder parameters, not just the scalar potential $g_{u}$ and non-Abelian
	vector potential $\{g_{A},g_{A3}\}$ parameters.
\end{thebibliography}
\end{document}